%% file: acl_lualatex.tex
\documentclass[11pt]{article}

\usepackage[preprint]{acl}

\usepackage{times}
\usepackage{latexsym}
\usepackage{amsmath}
\usepackage{amssymb}
\usepackage[T1]{fontenc}

\usepackage[utf8]{inputenc}

\usepackage{microtype}

\usepackage{inconsolata}

\usepackage{graphicx}
\usepackage[table]{xcolor}
\usepackage{multirow}
\usepackage{booktabs}
\usepackage{algorithm}
\usepackage{algorithmic}

\title{Learning User-Aware Recall:\\ Personalized Retrieval in Long-Term Conversational Memory}

\author{\mdseries
  ZhiShu Jiang\textsuperscript{1},
  Haibo Liu\textsuperscript{1},
  Xin Shen\textsuperscript{1,2},
  Guanqiang Qi\textsuperscript{1},
  Chenxi Miao\textsuperscript{1},
\\
  Weikang Li\textsuperscript{1,$\dagger$},
  Liwei Qian\textsuperscript{1},
  Xin Pei\textsuperscript{1},
  Jizhou Huang\textsuperscript{1}
\\
\\
  \textsuperscript{1}Baidu Inc. \quad \textsuperscript{2}The University of Queensland
\\
  {\small \texttt{jiangzhishu@bjtu.edu.cn, wavejkd@pku.edu.cn}}
\thanks{$\dagger$ Corresponding author}
}

\begin{document}
\maketitle
\input{0_abs}
\input{1_intro}

\input{2_related_work}
\input{3_method}
\input{4_experiment}
\input{5_conclusion}

\bibliography{custom}

\input{x_Appendix}


\end{document}

%% file: 0_abs.tex
\begin{abstract}
Long-term conversational agents are expected to remember past interactions, but memory is useful only when the right evidence is recalled for the right user. 
Existing memory-augmented LLM agents have made progress in building compact memory banks, yet retrieval is still often driven by query-centered similarity or fixed ranking rules, leaving user-conditioned relevance underexplored.
To address this gap, we propose \textbf{P}rofile-guided \textbf{P}ersonalized \textbf{R}etrieval \textbf{O}ptimization (\textbf{PPRO}), a retrieval-centric framework that makes memory retrieval both user-aware and optimizable.
PPRO builds episodic and semantic memory banks from dialogue histories and derives a user profile from accumulated memories.
The profile serves as an explicit personalized prior in memory ranking, allowing retrieval to account for stable user attributes, preferences, and relationships.
PPRO further trains a query rewriter with Group Relative Policy Optimization, using both evidence retrieval quality and downstream answer quality as feedback while keeping the memory banks and answer model fixed.
Experiments on LoCoMo and LongMemEval-S show consistent gains over training-free memory systems and training-based baselines.
Ablation studies further show that both profile-guided ranking and retrieval-oriented rewriting contribute substantially to performance, highlighting retrieval optimization as a key factor in personalized long-term memory use.
\end{abstract}

%% file: 1_intro.tex
\begin{figure}[ht!]
\centering
\includegraphics[width=0.48\textwidth]{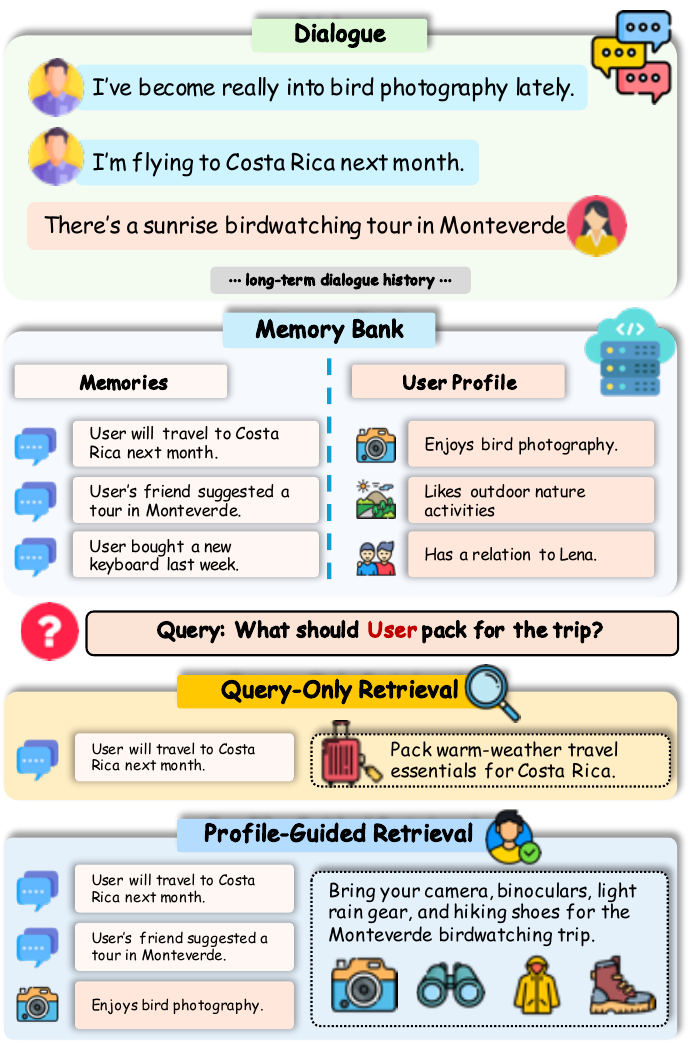}
\caption{
\textbf{Motivating example of profile-guided personalized memory retrieval.} 
Query-only retrieval may return surface-matching evidence, while profile-guided retrieval uses the user’s preferences, plans, and relationships to produce a more personalized answer.
}
\label{fig:teaser}
\vspace{-1.5em}
\end{figure}

\section{Introduction}

Large language model (LLM) agents are increasingly expected to support long-term personalized interaction, where useful knowledge must be accumulated across extended dialogue histories and reused in later conversations~\citep{maharana-etal-2024-evaluating,packer2023memgpt,wu2026true}.
Recent memory-augmented systems have made substantial progress in reducing the cost of long-context inputs by extracting, compressing, and organizing historical interactions into external memory banks~\citep{chhikara2025mem0,xu2025amem,liu2026simplemem,kang2025memoryos,lewis2020rag,shen2026duccae,liao2026stay}.
As these systems mature, a central challenge moves from memory construction alone to memory use at inference time: given a user query, the agent must retrieve evidence that is not only semantically relevant to the query, but also appropriate for the user behind the query~\cite{shen2021text}.
In long-term dialogue, different users accumulate distinct memories that reflect their relationships, topical interests, and life contexts~\cite{shen2021identifying,shen2021learning}.
The same query may therefore correspond to different information needs depending on who is asking.
As illustrated in Figure~\ref{fig:teaser}, personalized memory retrieval requires going beyond surface query matching: the retriever must identify evidence that is jointly relevant to the query and to the user’s long-term profile.

Existing memory systems have not fully addressed this user-conditioned retrieval problem.
Most work focuses on memory construction and management, including extraction, compression, hierarchical organization, graph-based indexing, and read-write operations~\citep{liu2026simplemem,xu2025amem,chhikara2025mem0}.
These methods improve the coverage and efficiency of stored memories, but retrieval is often treated as a fixed downstream step based on semantic matching or heuristic scoring~\citep{gutierrez2024hipporag,kim2025synapticrag,karpukhin2020dpr}.
Personalized dialogue systems, on the other hand, model user-specific information~\citep{zhu2025prime,liang2025rmm,zhang2018personachat,salemi2024lamp}, but usually focus on constructing personalized memory content or using profiles during generation rather than injecting user-level priors into retrieval ranking.
As a result, current memory pipelines may retrieve memories that match the query surface while failing to prioritize evidence that is more informative under the user’s stable attributes, preferences, and relationships.

To address these issues, we propose \textbf{P}rofile-guided \textbf{P}ersonalized \textbf{R}etrieval \textbf{O}ptimization (\textbf{PP RO}), a retrieval-centric framework for personalized long-term conversational memory.
PPRO first constructs a hierarchical memory representation from dialogue histories. 
It extracts episodic memories as fine-grained factual statements, aggregates related episodic memories into semantic memories, and summarizes stable user attributes, preferences, and relationships into a user profile. 
During inference, PPRO performs dual-path retrieval over both episodic and semantic memories. 
More importantly, the user profile is not only provided to the answer model as context, but also injected into the retrieval score as an embedding-level personalized prior. 
This design allows the retriever to rank memories by considering both query relevance and user-level relevance, making memory retrieval more sensitive to personalized information needs.

PPRO further optimizes retrieval through a trainable query rewriter.
Because retrieving discrete memory items is non-differentiable, gradients cannot flow from answer quality back to the retriever. To bypass this, we train a query rewriter with Group Relative Policy Optimization (GRPO)~\citep{shao2024deepseekmath,schulman2017ppo}: multiple rewritten queries are sampled and rewarded based on both evidence retrieval quality and final answer quality, while the memory banks, embedding model, and answer model remain frozen.


We evaluate PPRO on LoCoMo~\citep{maharana-etal-2024-evaluating} and LongMemEval-S~\citep{wu2024longmemeval}, two benchmarks for long-term conversational question answering.
PPRO consistently outperforms both training-free memory systems and training-based baselines across both benchmarks, and ablation studies confirm the contribution of user-profile guidance, dual-path retrieval, retrieval dimensions, and GRPO-based query rewriting.
The results show that personalized retrieval optimization is an effective complement to memory construction.
Overall, the contributions of our work are threefold:
\vspace{-0.5em}
\begin{itemize}
\setlength{\itemsep}{0pt}    
\setlength{\parsep}{0pt}     
\setlength{\parskip}{0pt}    
\item We formulate personalized memory retrieval as a key problem in long-term conversational agents, emphasizing that retrieved evidence should be relevant to both the input query and the target user.
\item We propose PPRO, combining profile-guided retrieval and GRPO-based query rewriting for personalized memory recall.
\item Experiments on LoCoMo and LongMemEval-S show consistent gains over both training-free and training-based baselines.
\end{itemize}
\vspace{-0.5em}

%% file: 2_related_work.tex
\section{Related Work}

\subsection{Long-Term Conversational Memory}

As the field progresses, research on long-term conversational memory has evolved from memory construction and organization, to retrieval mechanism design, and more recently to learning-based optimization~\citep{sumers2024coala,jiang2024omne}.

\noindent\textbf{Memory Construction and Organization.} Effective memory organization is fundamental to long-term conversational agents. MemGPT~\citep{packer2023memgpt} introduces OS-inspired virtual context management. A-Mem~\citep{xu2025amem} employs Zettelkasten-inspired dynamic indexing, Mem0~\citep{chhikara2025mem0} builds graph-based memory banks, and MemoryOS~\citep{kang2025memoryos} proposes a multi-level memory operating system. SimpleMem~\citep{liu2026simplemem} achieves strong performance through semantic lossless compression with multi-view indexing.

\noindent\textbf{Memory Retrieval Mechanisms.} Beyond better memory organization, how to accurately retrieve relevant memories given a query is equally critical. HippoRAG~\citep{gutierrez2024hipporag} draws from hippocampal indexing theory for multi-hop retrieval. THEANINE~\citep{bae2025theanine} retrieves via temporal-causal graph traversal, SynapticRAG~\citep{kim2025synapticrag} applies biologically-inspired synaptic propagation for temporal scoring, Zep~\citep{rasmussen2025zep} leverages temporal knowledge graphs, and CDMem~\citep{wang2025cdmem} applies context-dependent indexing conditioned on task state.

\noindent\textbf{RL-Based Memory Optimization.} More recently, reinforcement learning has been applied to optimize memory systems end-to-end. Memory-R1~\citep{yan2025memoryr1} trains policies for memory CRUD operations via GRPO. MEM1~\citep{zhou2025mem1} optimizes the think-search-answer loop, MemAgent~\citep{lu2025memagent} trains segmented reading for long-context compression, and Mem-T~\citep{memt2026} employs memory-of-thought augmented GRPO. However, RL in these methods targets memory management or reasoning without being tightly coupled with the retrieval process. Our work unifies reinforcement learning with user-aware personalized retrieval, optimizing the retrieval process with reward signals derived from both evidence retrieval quality and answer quality.

\subsection{User-Aware Retrieval}

User-aware retrieval focuses on leveraging user-specific information to improve retrieval relevance. In recommendation, recent work enhances user-aware retrieval through personalized representations and preference modeling~\citep{tan2024idgenrec,kong2024lora,chen2024sdpo,zhang2024gpg,jia2025learn}. In conversation, PRIME~\citep{zhu2025prime} maps cognitive dual-memory theory onto LLM personalization, RMM~\citep{liang2025rmm} introduces reflective memory management with RL-refined retrieval, InsideOut~\citep{zhao2026insideout} evolves user-centric core memory trees, and EMG-RAG~\citep{lee2024emgrag} crafts personalized agents through editable memory graphs. Our work injects user profile embeddings directly into the retrieval scoring function as a personalized ranking prior, making user-aware retrieval end-to-end optimizable.
    

%% file: 3_method.tex
\section{Methodology}

\begin{figure*}[t]
\centering
\includegraphics[width=0.95\textwidth]{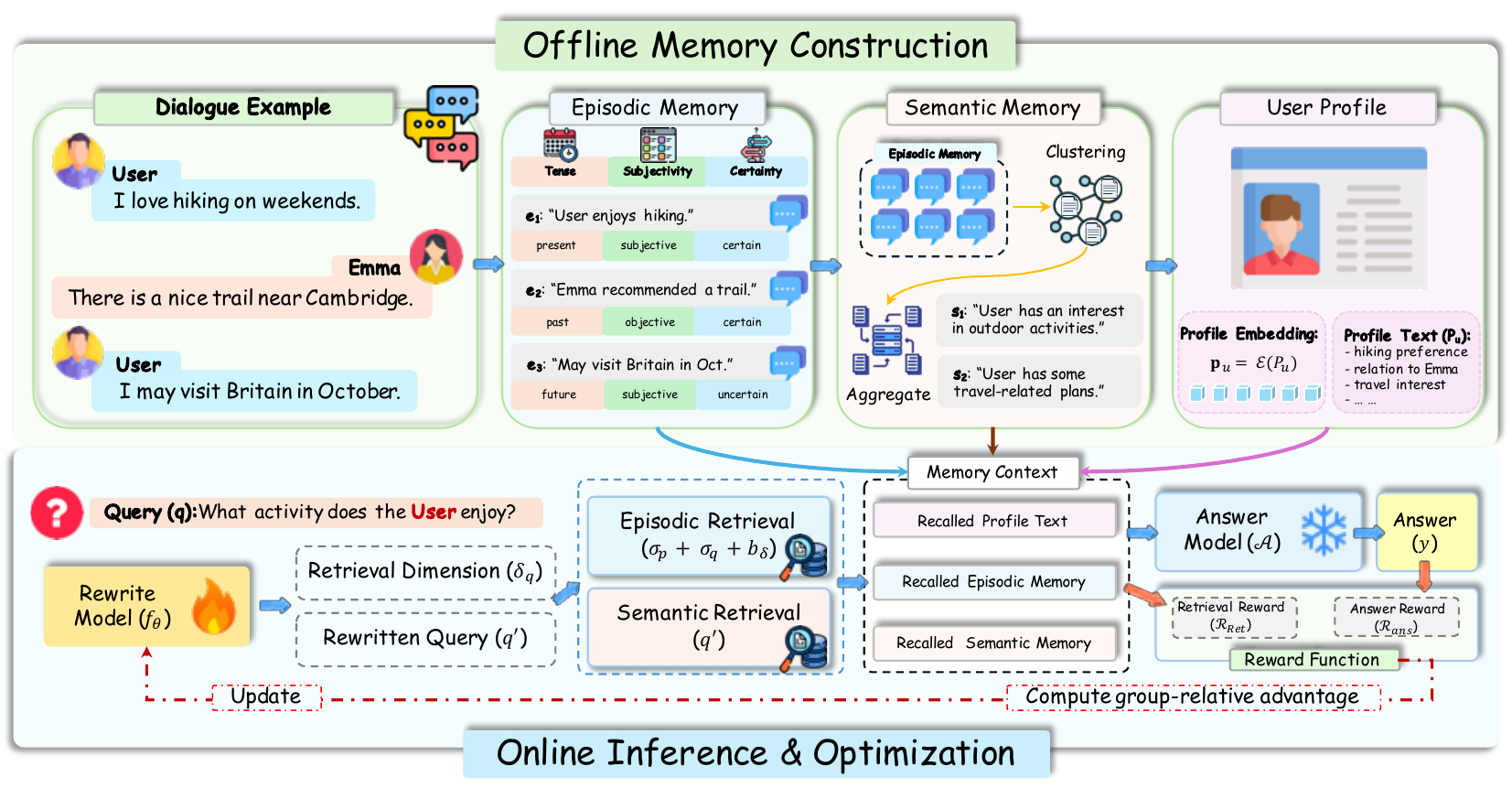}
\vspace{-1em}
\caption{Overview of the PPRO framework. The upper panel shows offline memory construction, while the lower panel shows online retrieval, answer generation, and GRPO optimization. PPRO builds episodic memories, semantic memories, and user profiles from dialogue histories, then uses profile-guided dual-path retrieval to recall memories. 
A GRPO-trained query rewriter further optimizes retrieval with evidence-level and answer-level feedback.}
\label{fig:framework}
\vspace{-1em}
\end{figure*}

We present Profile-guided Personalized Retrieval Optimization (\textbf{PPRO}), a retrieval-centric framework for personalized long-term conversational memory. 
As shown in Figure~\ref{fig:framework}, PPRO constructs episodic memories, semantic memories, and user profiles from dialogue histories, and then uses the profile as a personalized prior to guide dual-path retrieval. 
A GRPO-trained query rewriter further optimizes retrieval using evidence retrieval quality and answer quality as feedback.

\subsection{Problem Formulation}

Given a target user $u$, we assume access to the user's long-term dialogue history $D_u=\{d_1,\ldots,d_T\}$, where each dialogue $d_t$ contains a sequence of timestamped utterances from multiple speakers. 
Given a history-grounded query $q$, the task is to generate an answer $y$ that matches the reference answer $y^*$.

We focus on personalized memory retrieval as the key intermediate step.
PPRO represents $D_u$ with an episodic memory bank $\mathcal{M}^E_u$, a semantic memory bank $\mathcal{M}^S_u$, and a user profile $\mathcal{P}_u$.
At inference time, the retriever selects memories from the episodic and semantic memory banks that are both query-relevant and user-relevant. 
These retrieved memories, together with the user profile, are then provided to a frozen answer model $\mathcal{A}$ to generate the final answer.

\subsection{Offline Memory Construction}

The offline phase converts raw dialogue histories into persistent user-specific memory resources. 
For each target user, PPRO builds three complementary components: episodic memories for fine-grained factual evidence, semantic memories for cross-dialogue aggregation, and a user profile for stable personalized information.
We use the same instruction-following language model $\mathcal{G}$ across all offline construction stages, with task-specific prompts.

\noindent\textbf{Episodic Memory.}
For each dialogue $d_t$, $\mathcal{G}$ converts dialogue utterances into a set of atomic episodic memories. 
Each memory unit $e$ is a factual textual statement paired with retrieval dimensions:
\begin{equation}
\mathcal{M}_{u,t}^{E} = \mathcal{G}(d_t,u), \quad
e=(c,\delta)\in \mathcal{M}_{u,t}^{E},
\end{equation}
where $c$ is the textual content and $\delta$ denotes the retrieval dimensions attached to the memory, including tense, subjectivity, and certainty. 
These retrieval dimensions provide structured cues for consistent memory extraction and are later used as dimension-aware matching signals in retrieval scoring.
The full episodic memory bank for user $u$ is $\mathcal{M}^E_u = \bigcup_{t=1}^{T} \mathcal{M}^E_{u,t}$, providing fine-grained factual evidence for downstream retrieval.

\noindent\textbf{Semantic Memory.}
Episodic memories are atomic and dialogue-local, making it difficult to answer queries that require cross-dialogue synthesis. 
To capture higher-level patterns, PPRO introduces a semantic memory bank $\mathcal{M}_u^S$ by clustering episodic memories based on cosine similarity:
\begin{equation}
\mathrm{sim}(e_i,e_j)=\cos(\mathcal{E}(e_i),\mathcal{E}(e_j)),
\end{equation}
where $\mathcal{E}$ denotes the embedding model. Episodic memories whose similarity exceeds a threshold $\tau_c$ are grouped into clusters $C_g$. 
For each resulting cluster $C_g$, the language model $\mathcal{G}$ summarizes its episodic memories into a semantic memory unit:
\begin{equation}
s = \mathcal{G}(C_g), \quad s \in \mathcal{M}_u^S.
\end{equation}

\noindent\textbf{User Profile.}
While episodic memories preserve local facts and semantic memories capture cross-dialogue patterns, many personalized queries benefit from a high-level representation of a user's stable characteristics, including attributes, preferences, and relationships. 
Inspired by personalized recommendation in recommender systems~\cite{nguyen2026amem4rec}, we propose aggregating user-level profiles to enhance memory retrieval during the online stage. 
PPRO constructs a textual user profile $\mathcal{P}_u$ by summarizing the user's semantic memory bank:
\begin{equation}
\mathcal{P}_u = \mathcal{G}(\mathcal{M}^S_u), \quad \mathbf{p}_u = \mathcal{E}(\mathcal{P}_u),
\end{equation}
where $\mathcal{G}$ summarizes long-term preferences, occupations, relationships, habits, and recurring interests, and $\mathcal{E}$ is the shared embedding model. 
The profile is stored both as text $\mathcal{P}_u$ and as an embedding vector $\mathbf{p}_u$, which later serves as a personalized prior for online retrieval.

\subsection{Online Inference and Optimization}
At the online stage, PPRO first rewrites the input query into a retrieval-oriented form, then retrieves complementary evidence from episodic and semantic memory banks, and finally generates an answer from the assembled memory context.
Personalization is introduced by injecting the user profile as a prior into episodic memory scoring, while the query rewriter is further optimized with GRPO using retrieval
and answer-quality feedback.
The complete procedure is summarized in Algorithm~\ref{alg:ppero} in the Appendix.

\noindent\textbf{Retrieval-Oriented Query Rewriting.}
The rewriter model $f_\theta$ transforms the original query $q$, conditioned on dialogue context $C$, into a retrieval-oriented query $q'$ and retrieval dimensions $\delta_q$:
\begin{equation}
(q', \delta_q) = f_\theta(q, C).
\end{equation}
The rewritten query $q'$ is used for memory retrieval, while $\delta_q$ provides dimension-level signals that complement semantic similarity in episodic retrieval.

\noindent\textbf{Dual-path Memory Retrieval.}
PPRO retrieves from episodic and semantic memory banks in parallel. The episodic path
focuses on fine-grained factual evidence, while the semantic path provides compact
cross-dialogue summaries.

\textit{Episodic Retrieval.} PPRO scores each candidate $e$ by fusing query relevance with a profile-based personalized prior, augmented by a retrieval-dimension boost:
\begin{equation}
\sigma(e) = \lambda\, \sigma_q(e) + (1-\lambda)\, \sigma_p(e) + b_\delta(e),
\end{equation}
where $\sigma_q(e)=\cos(\mathcal{E}(q'), \mathcal{E}(e))$ measures query-memory relevance, $\sigma_p(e)=\cos(\mathbf{p}_u, \mathcal{E}(e))$ captures user-profile relevance via the profile embedding $\mathbf{p}_u$, and $b_\delta(e)$ is the number of matching retrieval dimensions between $\delta_q$ and $\delta_e$ multiplied by a fixed bonus (0.05 per match across three dimensions: tense, subjectivity, and certainty).
The hyperparameter $\lambda$ controls the trade-off between query-driven and profile-driven retrieval.

\textit{Semantic Retrieval.} In parallel, PPRO retrieves from $\mathcal{M}^S_u$ using the same rewritten query $q'$. 
Because semantic memories are already aggregated from user-specific episodic memories, they provide compact high-level context and are ranked directly by query similarity.
Semantic memories are therefore ranked by standard cosine similarity $\cos(\mathcal{E}(q'), \mathcal{E}(s))$.

\noindent\textbf{Answer Generation.}
The retrieved memories are concatenated into a unified memory context $\mathcal{C}_m = [\mathcal{P}_u; \mathcal{R}^E; \mathcal{R}^S]$, and the answer model $\mathcal{A}$ generates the answer:
\begin{equation}
y = \mathcal{A}(q, \mathcal{C}_m).
\end{equation}

\subsection{GRPO-based Optimization}
The quality of memory-augmented QA depends heavily on whether the rewritten query can retrieve the correct evidence.
Following recent work that uses multi-aspect feedback to train query rewriters~\cite{wang2024maferw,ma2023rewrite,wang2023query2doc}, we optimize the rewriter $f_\theta$ through the online inference pipeline using Group Relative Policy Optimization (GRPO)~\cite{shao2024deepseekmath}.
Instead of optimizing rewriting quality in isolation, $f_\theta$ is trained with feedback from both evidence retrieval quality and final answer quality.

\noindent\textbf{Policy and Environment.}
The rewriter $f_\theta$ is treated as the policy $\pi_\theta$. 
For each training data, the state is the original query $q$ with its dialogue context $C$, and the policy samples a group of $N$ actions, corresponding to rewritten queries $\{q'_1,\ldots,q'_N\}$ from $\pi_\theta(\cdot \mid q, C)$. 
For each sampled rewrite $q'_i$, PPRO executes the online inference pipeline to obtain retrieved episodic memories $\mathcal{R}^E_i$, semantic memories $\mathcal{R}^S_i$, and a generated answer $y_i$.

\noindent\textbf{Reward Design.}
The reward combines a retrieval reward and an answer reward:
\begin{equation}
R_i = \alpha \widetilde{R}_{ans,i} + (1-\alpha) \widetilde{R}_{ret,i},
\end{equation}
where $\alpha$ controls the trade-off between the two signals. 
The retrieval reward $R_{ret,i}$ is the F1 score between the source turns of retrieved episodic memories and ground-truth evidence annotations.
The answer reward $R_{ans,i} = \mathrm{BLEU\text{-}1}(y_i, y^*)$ measures unigram precision between the generated answer and the reference~\cite{papineni-etal-2002-bleu}.
Both rewards are normalized via EMA-based adaptive scaling before being combined. 

\noindent\textbf{Group-relative Advantage and Optimization.}
After executing the pipeline for each of the $N$ rewrites, the group-relative advantage is:
\begin{equation}
\hat{A}_i = \frac{R_i - \mathrm{mean}(\{R_j\}_{j=1}^{N})}{\mathrm{std}(\{R_j\}_{j=1}^{N}) + \epsilon}.
\end{equation}
This eliminates the need for a separately trained critic and directly compares alternative rewrites for the same query. The rewriter is then optimized with the standard GRPO clipped policy-gradient objective with a KL penalty (full formulation in \textit{Appendix}~\ref{sec:appendix_method_details}). 

%% file: 4_experiment.tex
\section{Experiment}

\begin{table*}[t]
\centering
\footnotesize
\setlength{\tabcolsep}{5pt}
\renewcommand{\arraystretch}{1.0}
\begin{tabular}{l|c|cc|cc|cc|cc|cc}
\toprule
\multirow{2}{*}{\textbf{Method}} & \multirow{2}{*}{\textbf{Base Model}} & \multicolumn{2}{c|}{\textbf{SingleHop}} & \multicolumn{2}{c|}{\textbf{Temporal}} & \multicolumn{2}{c|}{\textbf{OpenDomain}} & \multicolumn{2}{c|}{\textbf{MultiHop}} & \multicolumn{2}{c}{\textbf{Overall}} \\
 & & F1 & B-1 & F1 & B-1 & F1 & B-1 & F1 & B-1 & F1 & B-1 \\
\midrule
\multicolumn{12}{l}{\textit{Comparison with training-free baselines on the LoCoMo full benchmark}} \\
\midrule
Full-Context & GPT-4o & \textbf{61.56} & \textbf{54.19} & 9.09 & 5.78 & 16.47 & 14.80 & 28.00 & 18.47 & 29.49 & 22.14 \\
MemGPT & GPT-4o & 40.16 & 36.55 & 17.29 & 13.18 & 12.24 & 11.87 & 30.36 & 22.83 & 28.30 & 22.65 \\
A-Mem & GPT-4o & 44.43 & 38.97 & 39.41 & 31.23 & 17.10 & 15.84 & 32.86 & 23.76 & 35.36 & 27.61 \\
LightMem & GPT-4o & 33.76 & 28.02 & 36.53 & 29.12 & 13.38 & 11.54 & 28.15 & 21.83 & 30.00 & 23.84 \\
Mem0 & GPT-4o & 39.12 & 35.43 & 52.38 & \underline{44.15} & 17.73 & 15.92 & 35.13 & 27.56 & 38.37 & \underline{31.73} \\
SimpleMem & GPT-4o & 45.41 & 39.25 & \underline{56.71} & 20.57 & \underline{18.23} & \underline{16.34} & \underline{35.89} & \underline{32.83} & \underline{40.87} & 30.42 \\
\rowcolor{gray!10}
\textbf{PPRO (w/o GRPO)} & GPT-4o & \underline{45.67} & \underline{40.73} & \textbf{59.68} & \textbf{48.68} & \textbf{27.06} & \textbf{24.25} & \textbf{47.00} & \textbf{41.17} & \textbf{48.16} & \textbf{41.60} \\
\cmidrule{1-12}
Full-Context & Qwen2.5-3B & 7.03 & 5.69 & 3.11 & 2.71 & 4.55 & 5.97 & 4.61 & 4.29 & 4.74 & 4.32 \\
MemGPT & Qwen2.5-3B & 7.26 & 5.52 & 2.94 & 2.95 & 7.04 & 7.10 & 5.07 & 4.31 & 5.15 & 4.42 \\
A-Mem & Qwen2.5-3B & 17.23 & 13.12 & \underline{27.59} & \underline{25.07} & 7.12 & 7.28 & 12.57 & 9.01 & 16.21 & 13.00 \\
LightMem & Qwen2.5-3B & 18.28 & 15.24 & 6.92 & 4.56 & 8.06 & 7.23 & 16.43 & 11.54 & 14.27 & 10.49 \\
Mem0 & Qwen2.5-3B & 16.47 & 12.43 & 8.52 & 6.23 & 10.24 & 8.82 & 16.89 & 11.56 & 14.65 & 10.44 \\
SimpleMem & Qwen2.5-3B & \underline{20.90} & \underline{18.01} & 21.47 & 19.50 & \underline{12.52} & \underline{10.19} & \underline{17.03} & \underline{11.87} & \underline{18.38} & \underline{14.48} \\
\rowcolor{gray!10}
\textbf{PPRO (w/o GRPO)} & Qwen2.5-3B & \textbf{30.27} & \textbf{23.04} & \textbf{49.39} & \textbf{37.08} & \textbf{16.97} & \textbf{13.95} & \textbf{37.01} & \textbf{32.08} & \textbf{37.11} & \textbf{30.34} \\
\cmidrule{1-12}
Full-Context & Qwen3-8B & 14.50 & 11.20 & 6.80 & 5.50 & 10.10 & 8.80 & 13.50 & 9.20 & 12.07 & 8.77 \\
MemGPT & Qwen3-8B & 11.50 & 9.10 & 5.50 & 4.20 & 12.50 & 10.80 & 14.20 & 9.80 & 11.79 & 8.57 \\
A-Mem & Qwen3-8B & 26.80 & 21.50 & 22.50 & 18.20 & 13.20 & 10.50 & 20.50 & 13.80 & 21.62 & 15.92 \\
LightMem & Qwen3-8B & 29.48 & 23.83 & 26.78 & 21.52 & 14.12 & 11.24 & 18.53 & 14.23 & 21.98 & 17.32 \\
Mem0 & Qwen3-8B & 33.05 & 27.24 & 32.48 & 26.13 & 15.23 & 12.54 & 22.42 & 16.83 & 26.02 & 20.41 \\
SimpleMem & Qwen3-8B & \textbf{46.62} & \textbf{40.69} & \underline{42.85} & \underline{36.49} & \underline{15.35} & \underline{13.90} & \underline{28.97} & \underline{24.93} & \underline{34.25} & \underline{29.54} \\
\rowcolor{gray!10}
\textbf{PPRO (w/o GRPO)} & Qwen3-8B & \underline{38.44} & \underline{29.32} & \textbf{53.38} & \textbf{43.97} & \textbf{24.45} & \textbf{19.20} & \textbf{43.35} & \textbf{37.89} & \textbf{43.36} & \textbf{36.42} \\
\midrule
\multicolumn{12}{l}{\textit{Comparison with training-based baselines on the LoCoMo test split}} \\
\midrule
MEM1 & Qwen2.5-7B & 15.44 & 14.21 & 14.73 & 12.60 & 20.54 & 19.01 & 21.63 & 16.65 & 19.07 & 15.56 \\
MemAgent & Qwen2.5-7B & \underline{34.84} & \underline{25.02} & \underline{21.09} & \underline{14.91} & \underline{23.44} & \underline{21.73} & \textbf{46.64} & \textbf{43.52} & \underline{37.97} & \underline{33.11} \\
\rowcolor{gray!10}
\rowcolor{gray!10}
\textbf{PPRO (Ours)} & Qwen2.5-7B & \textbf{39.39} & \textbf{29.81} & \textbf{52.50} & \textbf{43.76} & \textbf{24.00} & \textbf{22.25} & \underline{43.34} & \underline{38.08} & \textbf{43.20} & \textbf{36.68} \\
\bottomrule
\end{tabular}
\vspace{-0.5em}
\caption{Token-level F1 and BLEU-1 results on LoCoMo. \textit{Upper}: comparison with training-free baselines on the full benchmark. \textit{Lower}: comparison with training-based baselines the test split. \textbf{Bold} = best, \underline{underline} = second best within each group.}
\label{tab:main_results}
\vspace{-1.5em}
\end{table*}

\subsection{Dataset}

We evaluate on two benchmarks that test long-term conversational memory: \textbf{LoCoMo}~\cite{maharana-etal-2024-evaluating} and \textbf{LongMemEval-S}~\cite{wu2024longmemeval}. LoCoMo contains multi-session dyadic conversations paired with QA annotations spanning four question categories: single-hop factual recall, temporal reasoning, open-domain, and multi-hop. Following the conversation-level split protocol of \citet{yan2025memoryr1}, we partition the ten conversations into a 1:1:8 train/validation/test split, yielding 152/81/1,307 QA pairs respectively (detailed statistics in Table~\ref{tab:locomo_split}). The training and validation sets are used exclusively for optimizing the query rewriter; all main results are reported on the held-out test set. LongMemEval-S~\cite{wu2024longmemeval} features exceptionally long interaction histories that require precise answer localization. We evaluate on five categories: \textit{Single-Session-User} (S-U), \textit{Single-Session-Preference} (S-P), \textit{Temporal-Reasoning} (Tmp.), \textit{Knowledge-Update} (K-U.), and \textit{Multi-Session} (Mlt.).

\subsection{Baselines}

We compare PPRO against two families of baselines (detailed descriptions in \textit{Appendix}~\ref{sec:appendix_baselines}). \textbf{Training-free baselines} include Full-Context, MemGPT~\citep{packer2023memgpt}, A-Mem~\citep{xu2025amem}, LightMem~\citep{fang2025lightmem}, Mem0~\citep{chhikara2025mem0}, and SimpleMem~\citep{liu2026simplemem}; we adopt results reported by \citet{liu2026simplemem} under identical evaluation metrics. \textbf{Training-based baselines} include MEM1~\citep{zhou2025mem1} and MemAgent~\citep{lu2025memagent}; we deploy their official Qwen2.5-7B checkpoints and evaluate on our test split.

\subsection{Evaluation Metrics}

We use three metrics across the two benchmarks. \textbf{Token-level F1}~\cite{rajpurkar-etal-2016-squad} measures the overlap between predicted and reference answer tokens; it is our primary metric on LoCoMo. \textbf{BLEU-1}~\cite{papineni-etal-2002-bleu} serves as a secondary metric that rewards precision and penalizes irrelevant tokens in the prediction. \textbf{LLM-as-Judge Accuracy} is used for LongMemEval-S: following \citet{liu2026simplemem}, we employ \texttt{gpt-4.1-mini} to produce binary \textsc{Correct}/\textsc{Wrong} labels and report accuracy. 

\begin{table*}[t]
\centering
\begin{minipage}[t]{0.48\textwidth}
\centering
\small
\setlength{\tabcolsep}{4pt}
\renewcommand{\arraystretch}{1.25}
\begin{tabular}{@{}lccccc|c@{}}
\toprule
\textbf{Method} & \textbf{Tmp.} & \textbf{Mlt.} & \textbf{K-U.} & \textbf{S-U} & \textbf{S-P} & \textbf{Ovr.} \\
\midrule
Full-context & 27.1 & 30.1 & 41.0 & 47.1 & 60.0 & 35.8 \\
Mem0 & 40.6 & 50.4 & 69.2 & \underline{87.1} & 63.5 & 57.4 \\
LightMem & \underline{85.7} & 47.4 & \textbf{92.3} & \textbf{88.6} & 76.7 & 75.2 \\
SimpleMem & 83.5 & \underline{60.9} & \underline{79.5} & 85.7 & \underline{76.7} & \underline{75.9} \\
\rowcolor{gray!10}
\textbf{PPRO (w/o GRPO)} & \textbf{89.3} & \textbf{74.4} & 80.8 & 81.4 & \textbf{80.0} & \textbf{81.5} \\
\bottomrule
\end{tabular}
\caption{LLM-as-Judge Accuracy on LongMemEval-S (weighted by category size). Baseline results are from~\citet{liu2026simplemem}.}
\label{tab:longmemeval_results}
\end{minipage}
\hfill
\begin{minipage}[t]{0.48\textwidth}
\centering
\small
\setlength{\tabcolsep}{13pt}
\renewcommand{\arraystretch}{1}
\begin{tabular}{@{}lcc@{}}
\toprule
\textbf{Setting} & \textbf{F1} & \textbf{B-1} \\
\midrule
Full Model & \textbf{43.20} & \textbf{36.68} \\
\midrule
w/o Episodic Memory & 34.34 & 28.98 \\
w/o Semantic Memory & 37.63 & 31.64 \\
w/o Profile Injection & 40.79 & 34.67 \\
w/o Retrieval Dimensions & 41.23 & 35.38 \\
w/o GRPO Training & 41.32 & 34.82 \\
\bottomrule
\end{tabular}
\caption{Ablation study on the LoCoMo test split with Qwen2.5-7B backbone. Per-category breakdown is in Table~\ref{tab:ablation_full}.}
\label{tab:ablation}
\end{minipage}
\vspace{-1.5em}
\end{table*}
\subsection{Experiment Setting}

To ensure fair comparison, we control the backbone of the answer model in PPRO to be identical to that of the baselines in each evaluation group. For LoCoMo, we evaluate with GPT-4o, Qwen2.5-3B, and Qwen3-8B as backbones for the comparison with training-free baselines, and Qwen2.5-7B for the comparison with training-based baselines. For LongMemEval-S, we use \texttt{gpt-4.1-mini} as the backbone following \citet{liu2026simplemem}. The rewriter model shares the same backbone as the answer model in each setting. Detailed hyperparameter settings, offline memory construction, and embedding models are provided in \textit{Appendix}~\ref{sec:appendix_settings}.

\begin{figure}[t]
\centering
\includegraphics[width=0.95\columnwidth]
{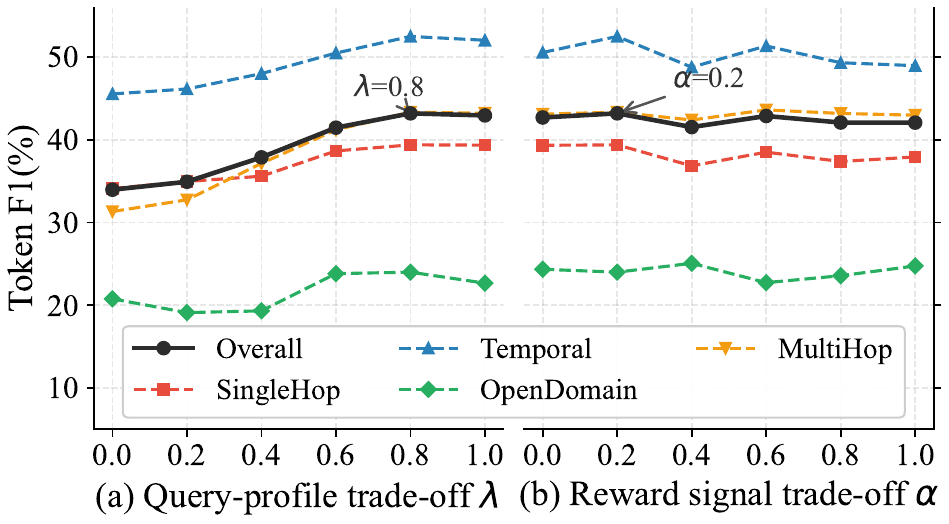}
\caption{Hyperparameter sensitivity on the LoCoMo test split. (a) Query-profile trade-off $\lambda$ (optimal at 0.8). (b) Reward signal trade-off $\alpha$ (optimal at 0.2).}
\label{fig:hyperparam}
\vspace{-1em}
\end{figure}

\subsection{Main Results}

\textbf{Results on LoCoMo.}
Table~\ref{tab:main_results} presents the main results on LoCoMo.
In the comparison with training-free baselines, PPRO achieves the best overall F1 across all three backbones, consistently outperforming the previous best SimpleMem by 7--19 points. The gains are particularly pronounced on Temporal and MultiHop categories, confirming the effectiveness of our retrieval dimensions and profile-guided personalization. On GPT-4o, Full-Context achieves the highest SingleHop F1 as its strong long-context capability suffices for single-hop recall. On Qwen3-8B, SimpleMem outperforms PPRO on SingleHop due to its adaptive retrieval size that avoids context overload for smaller models.

Compared to training-based baselines, PPRO achieves the best overall F1, outperforming MemAgent by more than 5 points, with the largest gains in Temporal and SingleHop. MemAgent achieves a higher MultiHop F1, likely because it is trained on HotpotQA which directly reinforces multi-hop evidence chaining.

\textbf{Results on LongMemEval-S.} Table~\ref{tab:longmemeval_results} shows that PPRO achieves the best overall accuracy, outperforming all baselines. The advantage is most evident on \textit{Temporal-Reasoning} and \textit{Multi-Session}, demonstrating that our hierarchical memory generalizes to longer interaction histories. PPRO also achieves the best \textit{Single-Session-Preference} accuracy, validating that the user profile effectively captures personalized preferences.

\subsection{Ablation Study}


We ablate each component by: (1) \textit{w/o Episodic Memory}: removing episodic memory; (2) \textit{w/o Semantic Memory}: removing semantic memory; (3) \textit{w/o Profile Injection}: disabling profile-guided retrieval and removing profile context from answer generation; (4) \textit{w/o Retrieval Dimensions}: disabling the retrieval dimensions in both rewriting and retrieval boosting; (5) \textit{w/o GRPO Training}: removing GRPO training for the query rewriter.

As shown in Table~\ref{tab:ablation}, all components contribute positively, with removing episodic memory causing the largest drop as it provides the primary evidence source. Removing semantic memory also leads to a substantial decline, confirming its complementary role in providing cross-dialogue aggregated context. Profile injection, retrieval dimensions, and GRPO training each provide consistent gains, demonstrating that they collectively enhance online retrieval quality through personalization, dimension matching, and learned query rewriting.
\begin{figure}[t]
\centering
\includegraphics[width=0.95\columnwidth]{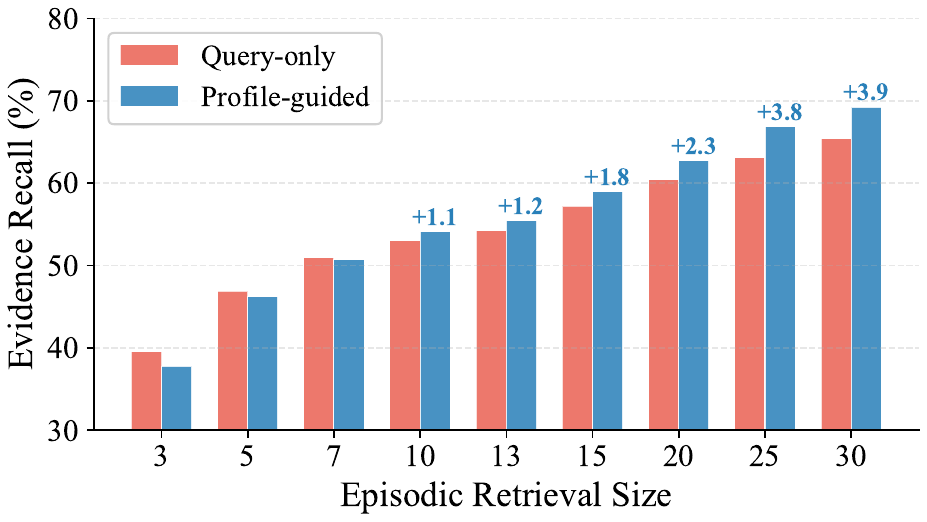}
\caption{Evidence Recall with and without profile-guided retrieval. Profile guidance yields increasing gains at larger retrieval sizes.}
\label{fig:evidence_recall}
\vspace{-1em}
\end{figure}

\begin{table*}[ht]
\centering
\scriptsize
\setlength{\tabcolsep}{2pt}
\renewcommand{\arraystretch}{1.15}
\begin{tabular}{p{1.5cm}| p{4.5cm}| p{5cm}| p{3.5cm}| c}
\toprule
\multicolumn{5}{l}{\textbf{Case Study: Resolving Temporally Confounded Charity Memories}} \\
\midrule
\multicolumn{1}{p{1.5cm}}{\textbf{Question}} 
& \multicolumn{4}{p{14cm}}{What was the main goal of the money raised from the charity tournament organized by John and his friends in May 2022?} \\

\multicolumn{1}{p{1.5cm}}{\textbf{Gold Answer}} 
& \multicolumn{4}{p{14cm}}{Raise money for a dog shelter.} \\

\multicolumn{1}{p{1.5cm}}{\textbf{Evidence}} 
& \multicolumn{4}{p{14cm}}{John: ``Our main goal was to raise money for a dog shelter, which is not far from the street where I live. And we did it!''} \\

\midrule
\textbf{Setting} & \textbf{Key Retrieved Signal} & \textbf{Model Prediction} & \textbf{Failure Mode} & \textbf{F1} \\
\midrule
No Memory 
& -- 
& No information available 
& No evidence 
& 0.00 \\

+ Episodic
& Children's hospital, Oct. 2022 
& To support a children's hospital 
& Temporal confusion 
& 0.18 \\

+ Semantic
& Dog shelter; secondary action: homeless 
& To raise money for charity and help the homeless 
& Primary/secondary goal confusion 
& 0.36 \\

+ User Profile 
& Charity/community-impact prior 
& To raise money for charity 
& Missing specific target 
& 0.40 \\

+ GRPO(Full) 
& May 2022 tournament matched to dog shelter 
& Raised money for a dog shelter near his street 
& Correct 
& 0.67 \\
\bottomrule
\end{tabular}
\caption{
Case study on temporally confounded memory retrieval.
}
\label{tab:case_study_charity}
\vspace{-2em}
\end{table*}

\subsection{Discussion}
\noindent\textbf{Hyperparameter Sensitivity.}
We analyze the sensitivity of PPRO to two key hyperparameters. 

\textit{Query-profile trade-off $\lambda$.} Figure~\ref{fig:hyperparam}(a) shows how the balance between query and profile embeddings affects retrieval quality. The query remains the dominant retrieval signal as performance generally increases with $\lambda$, yet injecting a small proportion of profile embedding ($\lambda=0.8$) achieves the best F1, surpassing query-only retrieval ($\lambda=1$).

\textit{Reward signal trade-off $\alpha$.} Figure~\ref{fig:hyperparam}(b) shows how the balance between retrieval and answer rewards affects downstream performance. The optimal $\alpha=0.2$ indicates that the retrieval reward should dominate the training signal, as it directly guides the rewriter toward better evidence coverage. The answer reward, while contributing a smaller proportion, still provides complementary guidance by ensuring the rewritten queries ultimately lead to better answer generation.


\begin{figure}[t]
\centering
\includegraphics[width=0.9\columnwidth]{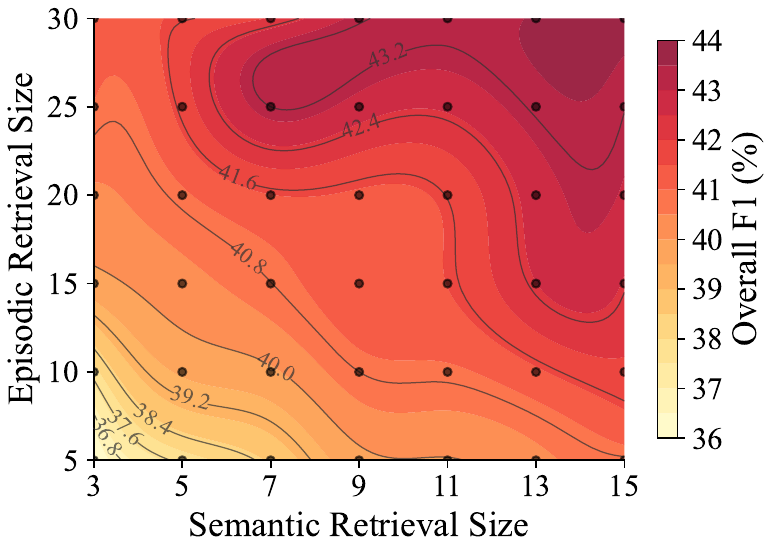}
\caption{Contour plot of Overall F1 (\%) over episodic and semantic memory retrieval sizes.}
\label{fig:topk_contour}
\vspace{-1.5em}
\end{figure}
\noindent\textbf{Retrieval Analysis.}
We further analyze retrieval quality from two perspectives.

\textit{Profile-Guided Retrieval Quality.} We measure Evidence Recall---the fraction of ground-truth evidence turns covered by retrieved episodic memories at varying retrieval sizes (formal definition in \textit{Appendix}~\ref{sec:appendix_evidence_recall}). Figure~\ref{fig:evidence_recall} compares profile-guided retrieval against query-only retrieval on episodic memory. At small sizes, the two methods perform comparably, but as the retrieval size grows, profile-guided retrieval increasingly outperforms the baseline. This confirms that the profile embedding acts as a relevance prior that helps rank relevant memories higher when the candidate pool is large.

\textit{Memory Layer Interaction.} Figure~\ref{fig:topk_contour} shows the joint effect of episodic and semantic memory retrieval sizes. Both layers contribute positively with clear diminishing returns. Performance changes more rapidly when increasing the semantic retrieval size in the low-to-moderate range, indicating that semantic memory provides greater marginal benefit per retrieved item due to its higher information density from cross-dialogue aggregation. Moderate retrieval sizes achieve near-optimal performance without excessive context length.

\subsection{Case Study}

To qualitatively illustrate how each component contributes to the final answer, Table~\ref{tab:case_study_charity} presents a representative example where PPRO progressively resolves a temporally confounded query. As components are added, each addresses a distinct failure mode, and only the full system produces the correct answer. This demonstrates that profile-guided retrieval and retrieval-oriented rewriting are complementary---the profile resolves user-level ambiguity, while the trained rewriter resolves temporal and factual specificity.

%% file: 5_conclusion.tex
\section{Conclusion}

We presented PPRO, a retrieval-centric framework that addresses user-agnostic ranking and task-agnostic optimization in memory-augmented conversational agents. PPRO constructs a hierarchical memory system and injects user profile embeddings as a personalized prior into retrieval scoring. A query rewriter trained with GRPO using evidence retrieval quality and answer quality as reward signals closes the optimization loop between query formulation and downstream performance without modifying memory banks or the answer model. Experiments on LoCoMo and LongMemEval-S demonstrate consistent improvements over both training-free and training-based baselines, confirming personalized retrieval optimization as a key factor in effective long-term memory use.

\section*{Limitations}

Although PPRO achieves effective personalized retrieval through hierarchical memory and profile-guided optimization, several limitations remain.

\noindent\textbf{Static Evaluation Setting.} PPRO is evaluated on static benchmarks where all dialogues are given in advance. We do not evaluate in streaming scenarios where memories must be incrementally updated during ongoing interaction.

\noindent\textbf{Supervised Reward Signal.} GRPO training relies on ground-truth evidence annotations for the retrieval reward. Extending the optimization to unsupervised settings where such annotations are unavailable remains to be explored.

\noindent\textbf{Single-User Scope.} The current framework models each user independently. In multi-party or community settings, shared context and inter-user relationships could further inform retrieval but are not captured by our per-user profile design.

\textit{Future Work:} We plan to extend PPRO to streaming incremental memory updates, explore self-supervised reward signals to remove annotation dependency, and investigate cross-user profile sharing for multi-party conversations.

%% file: x_Appendix.tex
\appendix
\section*{Appendix}
\noindent This appendix is organized as follows:

\begin{itemize}
\item Method details (Section~\ref{sec:appendix_method_details})
\item Experiment setup (Section~\ref{sec:appendix_experiment_setup})
\item Detailed ablation study (Section~\ref{sec:appendix_ablation})
\item Additional case studies (Section~\ref{sec:appendix_case_studies})
\item Prompts (Section~\ref{sec:appendix_prompts})
\item LLM usage statement (Section~\ref{LLM Usage Statement})
\end{itemize}

\section{Method Details}
\label{sec:appendix_method_details}

\subsection{Retrieval Reward}

The retrieval reward $R_{ret,i}$ measures how well the retrieved episodic memories cover the ground-truth evidence. Let $\mathcal{V}_q$ denote the annotated evidence turn set for query $q$, and let $\hat{\mathcal{V}}_i = \bigcup_{e \in \mathcal{R}^E_i} \mathrm{src}(e)$ be the set of source turns from the retrieved episodic memories $\mathcal{R}^E_i$. The retrieval reward is defined as:
\begin{equation}
R_{ret,i} = F_1(\hat{\mathcal{V}}_i,\, \mathcal{V}_q) = \frac{2P_i \mathrm{Rec}_i}{P_i + \mathrm{Rec}_i},
\end{equation}
where $P_i = |\hat{\mathcal{V}}_i \cap \mathcal{V}_q| / |\hat{\mathcal{V}}_i|$ and $\mathrm{Rec}_i = |\hat{\mathcal{V}}_i \cap \mathcal{V}_q| / |\mathcal{V}_q|$.

\subsection{EMA-based Reward Normalization}

Because the retrieval reward $R_{ret}$ and answer reward $R_{ans}$ can have different scales and learning dynamics, we apply EMA-based adaptive normalization before combining them:
\begin{equation}
\widetilde{R}_{*,i}=\frac{R_{*,i}}{\max(\mu_{*},\epsilon)},
\end{equation}
where $\mu_{*}$ denotes the exponential moving average of each respective reward stream ($* \in \{ret, ans\}$). This normalization prevents one signal from dominating merely due to scale differences.

\subsection{GRPO Objective}

The GRPO objective optimizes the rewriter with a clipped policy-gradient surrogate and a KL penalty against a frozen reference policy $\pi_{ref}$:
\begin{equation}
\begin{split}
\mathcal{L}_{GRPO}(\theta) = -\mathbb{E}_{i,t}\big[&\min\big(r_{i,t}\hat{A}_i,\; \mathrm{clip}(r_{i,t},1\!-\!\varepsilon, \\
1\!+\!\varepsilon)\hat{A}_i\big)
&- \mu D_{KL}\big],
\end{split}
\end{equation}
where $\mathbb{E}_{i,t}$ averages over all $N$ sampled rewrites and their token positions, and the importance ratio is
\begin{equation}
r_{i,t}=\frac{\pi_\theta(q'_{i,t}\mid q'_{i,<t},q,C)}{\pi_{\theta_{\mathrm{old}}}(q'_{i,t}\mid q'_{i,<t},q,C)}.
\end{equation}
The KL divergence is estimated token-wise following DeepSeekMath~\cite{shao2024deepseekmath}. The answer model, embedding model, and memory banks remain frozen; gradients are applied only to the rewriter.

\subsection{Retrieval Dimensions}
\label{sec:appendix_retrieval_dimensions}

Each episodic memory and each rewritten query is annotated with three retrieval dimensions:
\begin{itemize}
\item \textbf{Tense}: \textit{present}, \textit{past}, or \textit{future} — indicates the temporal orientation of the statement.
\item \textbf{Subjectivity}: \textit{objective} or \textit{subjective} — distinguishes factual statements from opinions or preferences.
\item \textbf{Certainty}: \textit{certain} or \textit{speculative} — differentiates confirmed facts from plans or hypotheticals.
\end{itemize}
During episodic retrieval, each dimension of the rewritten query $\delta_q$ is compared against the corresponding dimension of each candidate memory $\delta_e$. A fixed bonus of $0.05$ is added to the retrieval score for each matching dimension, yielding a maximum boost of $0.15$ when all three dimensions align.

\subsection{Evidence Recall}
\label{sec:appendix_evidence_recall}

Evidence Recall measures how well the retrieved memories cover the ground-truth evidence for a given query. Each query $q$ is annotated with a set of evidence turn identifiers $\mathcal{V}_q$. Each retrieved memory $e$ is associated with its source turn(s) $\mathrm{src}(e)$ via provenance tracking. Given the retrieved episodic memories $\mathcal{R}$ at a given retrieval size:
\begin{equation}
\mathrm{EvidenceRecall} = \frac{|\mathcal{V}_q \cap \bigcup_{e \in \mathcal{R}} \mathrm{src}(e)|}{|\mathcal{V}_q|}
\end{equation}

\subsection{Algorithm}

\begin{algorithm}[h]
\caption{PPRO: Inference and Training}
\label{alg:ppero}
\begin{algorithmic}[1]
\REQUIRE Query $q$, dialogue context $C$, user $u$, memory banks $\mathcal{M}^E_u$, $\mathcal{M}^S_u$, profile $\mathcal{P}_u$
\ENSURE Answer $y$
\STATE \textit{// Phase 1: Online Inference}
\STATE $(q', \delta_q) \leftarrow f_\theta(q, C)$ \COMMENT{Query rewriting}
\STATE $\mathbf{p}_u \leftarrow \mathcal{E}(\mathcal{P}_u)$ \COMMENT{Profile embedding}
\STATE $\mathcal{R}^E \leftarrow \mathrm{Retrieve}(\mathcal{M}^E_u, q', \mathbf{p}_u, \delta_q, \lambda)$ \COMMENT{Episodic}
\STATE $\mathcal{R}^S \leftarrow \mathrm{Retrieve}(\mathcal{M}^S_u, q')$ \COMMENT{Semantic}
\STATE $\mathcal{C}_m \leftarrow [\mathcal{P}_u;\; \mathcal{R}^E;\; \mathcal{R}^S]$
\STATE $y \leftarrow \mathcal{A}(q, \mathcal{C}_m)$ \COMMENT{Answer generation}
\STATE
\STATE \textit{// Phase 2: GRPO Training}
\STATE Sample $N$ rewrites $\{q'_i\}_{i=1}^N \sim \pi_\theta(\cdot \mid q, C)$
\FOR{$i = 1$ to $N$}
  \STATE Execute Phase 1 with $q'_i$ to obtain $y_i$
  \STATE Compute $R_{ret,i}$ and $R_{ans,i}$
  \STATE $R_i \leftarrow \alpha \widetilde{R}_{ans,i} + (1-\alpha) \widetilde{R}_{ret,i}$
\ENDFOR
\STATE $\hat{A}_i \leftarrow (R_i - \mathrm{mean}(\{R_j\})) / (\mathrm{std}(\{R_j\}) + \epsilon)$
\STATE Update $\theta$ via $\mathcal{L}_{GRPO}(\theta)$
\end{algorithmic}
\end{algorithm}

\section{Experiment Setup}
\label{sec:appendix_experiment_setup}

\subsection{Dataset Split Statistics}
\label{sec:appendix_data}

\textbf{LoCoMo.} The LoCoMo dataset~\citep{maharana-etal-2024-evaluating} contains 10 multi-session dyadic conversations, each spanning 19--35 sessions between two speakers. Questions are annotated into five categories: SingleHop (single-session factual recall), Temporal (temporal reasoning requiring date/order understanding), OpenDomain (requiring external or commonsense knowledge), MultiHop (cross-session evidence synthesis), and Adversarial (unanswerable questions). Following the conversation-level split protocol of \citet{yan2025memoryr1}, we assign conversations to train/validation/test splits at the conversation level (1:1:8 ratio) to prevent information leakage across splits. Category 5 (Adversarial) questions are excluded from all evaluations, as their gold answer is uniformly ``unanswerable'' and they do not test memory retrieval capability. The resulting statistics are shown in Table~\ref{tab:locomo_split}.

\textbf{LongMemEval-S.} LongMemEval~\citep{wu2024longmemeval} provides exceptionally long interaction histories (up to 500+ turns) and tests fine-grained answer localization. We use its short-form QA subset (LongMemEval-S) which contains 5 question categories: \textit{Single-Session-User} (user-stated facts within one session), \textit{Single-Session-Preference} (user preferences expressed in one session), \textit{Temporal-Reasoning} (temporal ordering and date-based questions), \textit{Knowledge-Update} (facts that evolve over time), and \textit{Multi-Session} (evidence scattered across multiple sessions). We evaluate on this benchmark in a zero-shot manner without any fine-tuning on its data.

\subsection{Evaluation Metrics}
\label{sec:appendix_metrics}

\textbf{Token-level F1.} Let $\mathcal{T}_p$ and $\mathcal{T}_r$ denote the multisets of whitespace-tokenized tokens in the prediction and reference respectively. Precision is $|\mathcal{T}_p \cap \mathcal{T}_r|/|\mathcal{T}_p|$, recall is $|\mathcal{T}_p \cap \mathcal{T}_r|/|\mathcal{T}_r|$, and F1 is their harmonic mean.

\textbf{BLEU-1.} Unigram BLEU with a brevity penalty, following \citet{papineni-etal-2002-bleu}. It complements Token-level F1 by rewarding precision and penalizing irrelevant tokens in the prediction.

\textbf{LLM-as-Judge Accuracy.} For LongMemEval-S, \texttt{gpt-4.1-mini} evaluates each predicted answer against the ground-truth reference, producing a binary \textsc{Correct}/\textsc{Wrong} label based on semantic equivalence and temporal consistency. Accuracy is the percentage of \textsc{Correct} labels.

\textbf{Implementation details.} Tokenization and BLEU-1 computation use NLTK 3.8~\citep{bird2009nltk} with \texttt{word\_tokenize} and \texttt{sentence\_bleu} (smoothing method 1). The BLEU-1 reward is computed with NLTK's \texttt{sentence\_bleu} (smoothing method 1). Sentence embeddings use \texttt{bge-large-en}~\citep{xiao2024cpack} via the FlagEmbedding library. GRPO training is implemented with the VERL framework~\citep{sheng2024verl}.

\subsection{Hyperparameter Settings}
\label{sec:appendix_settings}

\textbf{Retrieval.} We use \texttt{bge-large-en} as the embedding model. Episodic retrieval top-$K$=25, semantic retrieval top-$K$=7, similarity threshold 0.6, query-profile mixing weight $\lambda$=0.8, and dimension boost $b_\delta$=0.05.

\textbf{Offline memory construction.} All offline stages (episodic extraction, semantic aggregation, and profile generation) use DeepSeek-V3.2~\citep{deepseekai2025deepseekv32} as the shared instruction-following model $\mathcal{G}$. The semantic memory clustering threshold is $\tau_c = 0.85$.

\textbf{GRPO training.} We use the VERL framework with 2$\times$ NVIDIA A800 80GB GPUs. Learning rate $1\times10^{-6}$, batch size 64, 30 epochs, group size $N$=8. Clip ratio $\varepsilon$=0.2, KL coefficient $\mu$=0.001, entropy coefficient 0.001, weight decay 0.01, gradient clipping 1.0. Answer reward weight $\alpha$=0.2, retrieval reward weight $(1-\alpha)$=0.8. Loss aggregation uses token-mean mode. We validate before training and select the best checkpoint based on validation F1.

\textbf{Computational resources.} The rewriter model shares the same backbone as the answer model in each evaluation setting. The embedding model is \texttt{bge-large-en} (335M parameters). Offline memory construction uses \texttt{qwen3-8b} (8B parameters). GRPO training is conducted on 2$\times$ NVIDIA A800 80GB GPUs; each training run takes approximately 15 hours ($\sim$30 GPU hours in total).

\subsection{Baseline Descriptions}
\label{sec:appendix_baselines}

\textbf{Training-free baselines.}
\textsc{Full-Context} feeds the complete dialogue history to the answer model without explicit memory construction. \textsc{MemGPT}~\citep{packer2023memgpt} manages memory through an operating-system-inspired virtual context mechanism. \textsc{A-Mem}~\citep{xu2025amem} stores memories as structured notes connected through dynamic indexing and memory evolution. \textsc{LightMem}~\citep{fang2025lightmem} separates lightweight online memory use from offline consolidation. \textsc{Mem0}~\citep{chhikara2025mem0} dynamically extracts and consolidates salient information with graph-based memory representations. \textsc{SimpleMem}~\citep{liu2026simplemem} applies semantic structured compression, online synthesis, and intent-aware retrieval planning.

\textbf{Training-based baselines.}
\textsc{MEM1}~\citep{zhou2025mem1} learns to maintain a compact internal memory state via PPO-trained iterative think-search-answer loops. \textsc{MemAgent}~\citep{lu2025memagent} uses DAPO to train a recurrent memory compression agent that processes long contexts through sequential chunk-level updates. Both methods are trained on HotpotQA.

\section{Detailed Ablation Study}
\label{sec:appendix_ablation}

Table~\ref{tab:ablation_full} presents the per-category breakdown of the ablation study. The trends are consistent with the overall results in the main text. Removing episodic memory causes the largest degradation, especially on MultiHop ($-12.23$ F1) where multiple fine-grained facts must be combined. Removing semantic memory impacts SingleHop and Temporal most, as aggregated timelines and counts directly serve these question types. Profile injection contributes most to OpenDomain ($-4.85$ F1), where user-specific relational context is needed for disambiguation. Retrieval dimensions show the largest impact on Temporal ($-4.88$ F1), confirming that tense matching effectively guides temporal retrieval. GRPO training benefits OpenDomain and Temporal most, as these categories require more specific query reformulation.

\section{Additional Case Studies}
\label{sec:appendix_case_studies}

Tables~\ref{tab:case_study_jewelry} and~\ref{tab:case_study_cooking} present two additional case studies following the same progressive format as Table~\ref{tab:case_study_charity} in the main text. Each table shows how incrementally adding PPRO components resolves a distinct retrieval challenge.

\textbf{Case 2: Distinguishing Motivation from Outcome} (Table~\ref{tab:case_study_jewelry}). The question asks \textit{why} Audrey makes recycled jewelry, but the memory bank contains both the stated motivation (``creativity and sustainability'') and a downstream outcome (``donate profits to an animal shelter''). Without the full system, retrieval conflates these two aspects. Episodic retrieval alone surfaces the more salient outcome memory (donation), while semantic memory partially captures sustainability but merges it with other environmental actions. The GRPO-trained rewriter learns to reformulate the query to specifically target stated motivations rather than behavioral consequences, surfacing the exact ``creativity and sustainability'' memory.

\textbf{Case 3: Stated Reason vs.\ Inferred Motivation} (Table~\ref{tab:case_study_cooking}). James explicitly stated he ``wanted to learn something new,'' but his memory bank also contains cooking-related social events (cooking for friends) and self-improvement themes. The challenge is distinguishing what the user \textit{actually said} from what could be \textit{inferred}. Episodic retrieval without GRPO surfaces socially-oriented cooking memories; semantic memory picks up the self-improvement theme but drifts toward inference (``challenge himself''). The profile further reinforces self-improvement but introduces paraphrase drift. Only the GRPO-trained rewriter generates a query specific enough to retrieve the verbatim stated reason, demonstrating that retrieval-oriented optimization helps the system distinguish explicit statements from plausible inferences.

\section{Prompts}
\label{sec:appendix_prompts}

\begin{figure*}[t]
\centering
\begin{minipage}[t]{0.5\textwidth}
  \centering
  \includegraphics[width=\textwidth]{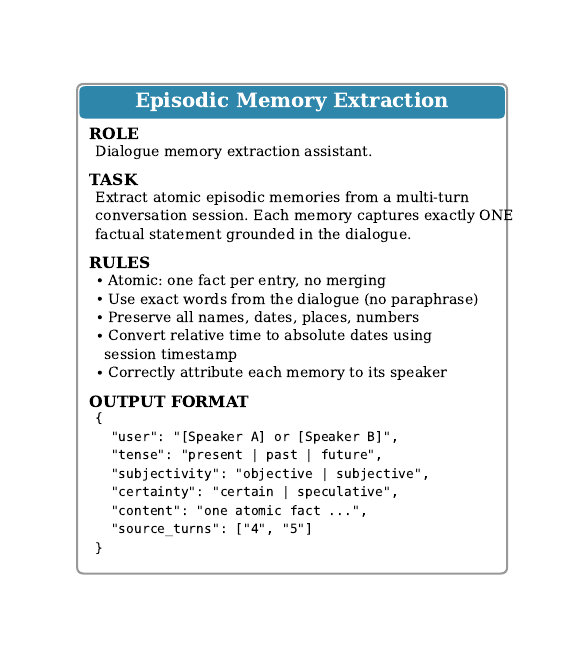}
  \caption{Prompt for episodic memory extraction.}
  \label{fig:prompt_episodic}
\end{minipage}\hfill
\begin{minipage}[t]{0.5\textwidth}
  \centering
  \includegraphics[width=\textwidth]{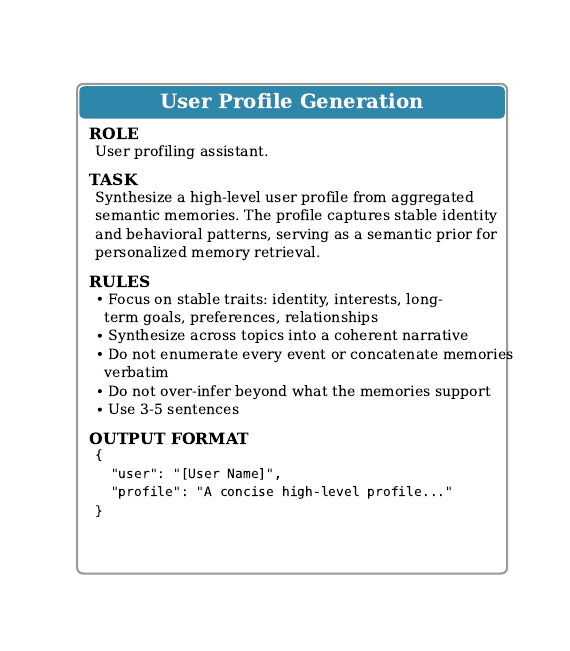}
  \caption{Prompt for user profile generation.}
  \label{fig:prompt_profile}
\end{minipage}

\vspace{1em}

\begin{minipage}[t]{0.5\textwidth}
  \centering
  \includegraphics[width=\textwidth]{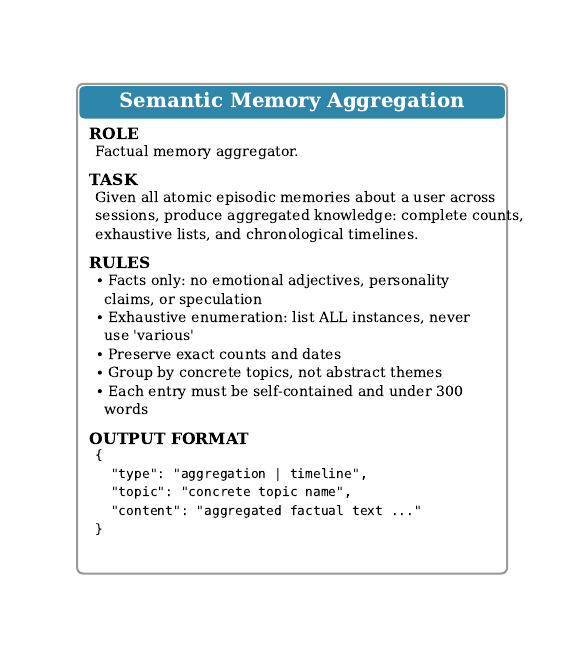}
  \caption{Prompt for semantic memory aggregation.}
  \label{fig:prompt_semantic}
\end{minipage}\hfill
\begin{minipage}[t]{0.5\textwidth}
  \centering
  \includegraphics[width=\textwidth]{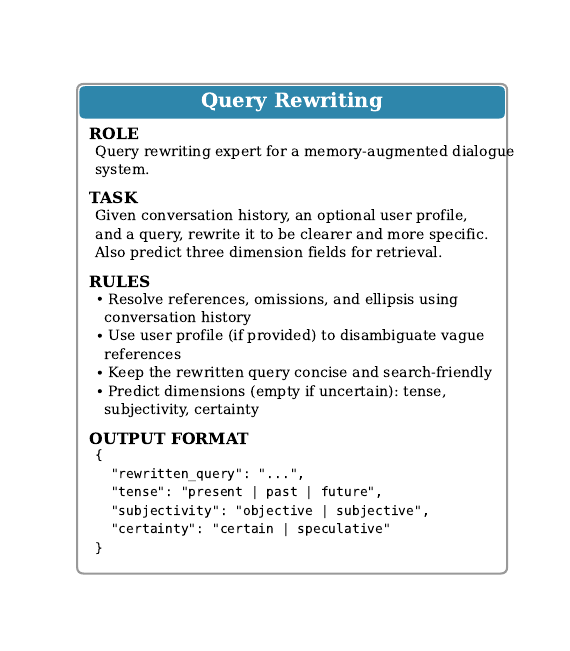}
  \caption{Prompt for profile-aware query rewriting.}
  \label{fig:prompt_rewrite}
\end{minipage}
\end{figure*}

Figures~\ref{fig:prompt_episodic}--\ref{fig:prompt_rewrite} show the simplified prompts used in the offline memory construction stages and the online query rewriting stage.

\section{LLM Usage Statement}
\label{LLM Usage Statement}
Large Language Models (LLMs) such as ChatGPT are used as general-purpose tools to improve readability and clarity of the manuscript, e.g., for grammar checking, LaTeX formatting, and restructuring sentences. 
No parts of the research idea, dataset design, or experimental results are generated or influenced by LLMs. 
All technical contributions and conclusions are solely those of the authors.

\begin{table*}[t]
\centering
\small
\setlength{\tabcolsep}{5pt}
\begin{tabular}{lccccccc}
\toprule
\textbf{Split} & \textbf{Conversations} & \textbf{Questions} & \textbf{SingleHop} & \textbf{Temporal} & \textbf{OpenDomain} & \textbf{MultiHop} & \textbf{Avg.\ Evidence} \\
\midrule
Train & 1 (conv-26) & 152 & 32 & 37 & 13 & 70 & 1.33 \\
Valid & 1 (conv-30) & 81 & 11 & 26 & 0 & 44 & 1.31 \\
Test & 8 (conv-41--50) & 1307 & 239 & 258 & 83 & 727 & 1.57 \\
\midrule
Total & 10 & 1540 & 282 & 321 & 96 & 841 & 1.52 \\
\bottomrule
\end{tabular}
\caption{LoCoMo data split statistics following the conversation-level protocol of \citet{yan2025memoryr1}. Category 5 (adversarial) questions are excluded. ``Avg.\ Evidence'' denotes the mean number of evidence turns per question. The training and validation sets are used exclusively for GRPO optimization of the query rewriter.}
\label{tab:locomo_split}
\end{table*}

\begin{table*}[t]
\centering
\small
\renewcommand{\arraystretch}{1.1}
\begin{tabular}{l|cc|cc|cc|cc|cc}
\toprule
\multirow{2}{*}{\textbf{Setting}} & \multicolumn{2}{c|}{\textbf{SingleHop}} & \multicolumn{2}{c|}{\textbf{Temporal}} & \multicolumn{2}{c|}{\textbf{OpenDomain}} & \multicolumn{2}{c|}{\textbf{MultiHop}} & \multicolumn{2}{c}{\textbf{Overall}} \\
 & F1 & B-1 & F1 & B-1 & F1 & B-1 & F1 & B-1 & F1 & B-1 \\
\midrule
Full Model & \textbf{39.39} & \textbf{29.81} & \textbf{52.50} & \textbf{43.76} & \textbf{24.00} & \textbf{22.25} & \textbf{43.34} & \textbf{38.08} & \textbf{43.20} & \textbf{36.68} \\
\midrule
w/o Episodic Memory & 34.00 & 25.98 & 47.74 & 39.78 & 22.04 & 18.67 & 31.11 & 27.30 & 34.34 & 28.98 \\
w/o Semantic Memory & 28.80 & 19.89 & 43.96 & 36.17 & 19.19 & 14.61 & 40.39 & 35.84 & 37.63 & 31.64 \\
w/o Profile Injection & 35.51 & 27.26 & 50.14 & 41.62 & 19.15 & 15.98 & 41.68 & 36.78 & 40.79 & 34.67 \\
w/o Retrieval Dimensions & 37.99 & 28.62 & 47.62 & 39.97 & 23.10 & 20.94 & 42.09 & 37.62 & 41.23 & 35.38 \\
w/o GRPO Training & 38.50 & 28.20 & 47.78 & 39.52 & 20.26 & 17.27 & 42.36 & 37.33 & 41.32 & 34.82 \\
\bottomrule
\end{tabular}
\caption{Full ablation study with per-category breakdown on the LoCoMo test split with Qwen2.5-7B backbone.}
\label{tab:ablation_full}
\end{table*}

\begin{table*}[t]
\centering
\scriptsize
\setlength{\tabcolsep}{2pt}
\renewcommand{\arraystretch}{1.15}
\begin{tabular}{p{1.5cm}| p{4.5cm}| p{5cm}| p{3.5cm}| c}
\toprule
\multicolumn{5}{l}{\textbf{Case Study: Distinguishing Motivation from Outcome}} \\
\midrule
\multicolumn{1}{p{1.5cm}}{\textbf{Question}}
& \multicolumn{4}{p{14cm}}{Why does Audrey make jewelry out of recycled objects?} \\

\multicolumn{1}{p{1.5cm}}{\textbf{Gold Answer}}
& \multicolumn{4}{p{14cm}}{To show love for creativity and sustainability.} \\

\multicolumn{1}{p{1.5cm}}{\textbf{Evidence}}
& \multicolumn{4}{p{14cm}}{Audrey: ``Oh yes! I love making jewelry out of recycled stuff. It's a great way to show my love of creativity and sustainability.''} \\
\midrule
\textbf{Setting} & \textbf{Key Retrieved Signal} & \textbf{Model Prediction} & \textbf{Failure Mode} & \textbf{F1} \\
\midrule
No Memory
& --
& No information available
& No evidence
& 0.00 \\

+ Episodic
& Donation profits; animal shelter
& To donate the profits to an animal shelter
& Wrong aspect
& 0.12 \\

+ Semantic
& Waste reduction; recycled materials
& To reduce waste and support a good cause
& Partial motivation
& 0.24 \\

+ User Profile
& Sustainability prior boosted
& To make a difference by reducing waste
& Incomplete dual framing
& 0.20 \\

+ GRPO(Full)
& ``creativity and sustainability'' memory surfaced
& To show her love of creativity and sustainability
& Correct
& 0.80 \\
\bottomrule
\end{tabular}
\caption{Case study: distinguishing motivation from outcome.}
\label{tab:case_study_jewelry}
\end{table*}

\begin{table*}[t]
\centering
\scriptsize
\setlength{\tabcolsep}{2pt}
\renewcommand{\arraystretch}{1.15}
\begin{tabular}{p{1.5cm}| p{4.5cm}| p{5cm}| p{3.5cm}| c}
\toprule
\multicolumn{5}{l}{\textbf{Case Study: Stated Reason vs.\ Inferred Motivation}} \\
\midrule
\multicolumn{1}{p{1.5cm}}{\textbf{Question}}
& \multicolumn{4}{p{14cm}}{Why did James sign up for a cooking class?} \\

\multicolumn{1}{p{1.5cm}}{\textbf{Gold Answer}}
& \multicolumn{4}{p{14cm}}{He wanted to learn something new.} \\

\multicolumn{1}{p{1.5cm}}{\textbf{Evidence}}
& \multicolumn{4}{p{14cm}}{James: ``I never liked cooking, but I felt that I wanted to learn something new.''} \\
\midrule
\textbf{Setting} & \textbf{Key Retrieved Signal} & \textbf{Model Prediction} & \textbf{Failure Mode} & \textbf{F1} \\
\midrule
No Memory
& --
& No information available
& No evidence
& 0.00 \\

+ Episodic
& Cooking events; social goal
& To cook for his friend John
& Wrong aspect
& 0.18 \\

+ Semantic
& ``Never liked cooking'' contrast
& Because he wanted to challenge himself
& Inferred not stated
& 0.29 \\

+ User Profile
& ``Improve himself'' boosted
& To improve himself by trying something new
& Paraphrase drift
& 0.50 \\

+ GRPO(Full)
& ``wanted to learn something new'' surfaced
& He felt he wanted to learn something new
& Correct
& 0.92 \\
\bottomrule
\end{tabular}
\caption{Case study: stated reason vs.\ inferred motivation.}
\label{tab:case_study_cooking}
\end{table*}

\clearpage

%% file: custom.bib
@article{nguyen2026amem4rec,
    title = "{AMEM4Rec}: Leveraging Cross-User Similarity for Memory Evolution in Agentic {LLM} Recommenders",
    author = "Nguyen, Minh-Duc and
      Kieu, Hai-Dang and
      Le, Dung D.",
    journal = "arXiv preprint arXiv:2602.08837",
    year = "2026"
}

@article{zhao2026insideout,
    title = "Inside Out: Evolving User-Centric Core Memory Trees for Long-Term Personalized Dialogue Systems",
    author = "Zhao, Jihao and
      Chen, Ding and
      Fan, Zhaoxin and
      Xu, Kerun and
      Hu, Mengting and
      Tang, Bo and
      Xiong, Feiyu and
      Li, Zhiyu",
    journal = "arXiv preprint arXiv:2601.05171",
    year = "2026"
}

@inproceedings{maharana-etal-2024-evaluating,
    title = "Evaluating Very Long-Term Conversational Memory of {LLM} Agents",
    author = "Maharana, Adyasha and
      Lee, Dong-Ho and
      Tulyakov, Sergey and
      Bansal, Mohit and
      Barbieri, Francesco and
      Fang, Yuwei",
    editor = "Ku, Lun-Wei and
      Martins, Andre and
      Srikumar, Vivek",
    booktitle = "Proceedings of the 62nd Annual Meeting of the Association for Computational Linguistics (Volume 1: Long Papers)",
    month = aug,
    year = "2024",
    address = "Bangkok, Thailand",
    publisher = "Association for Computational Linguistics",
    url = "https://aclanthology.org/2024.acl-long.747/",
    doi = "10.18653/v1/2024.acl-long.747",
    pages = "13851--13870"
}

@article{yan2025memoryr1,
    title = "{Memory-R1}: Enhancing Large Language Model Agents to Manage and Utilize Memories via Reinforcement Learning",
    author = "Yan, Sikuan and
      Yang, Xiufeng and
      Huang, Zuchao and
      Nie, Ercong and
      Ding, Zifeng and
      Li, Zonggen and
      Ma, Xiaowen and
      Bi, Jinhe and
      Kersting, Kristian and
      Pan, Jeff Z. and
      Sch{\"u}tze, Hinrich and
      Tresp, Volker and
      Ma, Yunpu",
    journal = "arXiv preprint arXiv:2508.19828",
    year = "2025",
    url = "https://arxiv.org/abs/2508.19828"
}

@article{shao2024deepseekmath,
    title = "{DeepSeekMath}: Pushing the Limits of Mathematical Reasoning in Open Language Models",
    author = "Shao, Zhihong and
      Wang, Peiyi and
      Zhu, Qihao and
      Xu, Runxin and
      Song, Junxiao and
      Bi, Xiao and
      Zhang, Haowei and
      Zhang, Mingchuan and
      Li, Y. K. and
      Wu, Y. and
      Guo, Daya",
    journal = "arXiv preprint arXiv:2402.03300",
    year = "2024",
    url = "https://arxiv.org/abs/2402.03300"
}

@article{wang2024maferw,
    title = "{MaFeRw}: Query Rewriting with Multi-Aspect Feedbacks for Retrieval-Augmented Large Language Models",
    author = "Wang, Yujing and
      Zhang, Hainan and
      Pang, Liang and
      Guo, Binghui and
      Zheng, Hongwei and
      Zheng, Zhiming",
    journal = "arXiv preprint arXiv:2408.17072",
    year = "2024",
    url = "https://arxiv.org/abs/2408.17072"
}

@article{chhikara2025mem0,
    title = "{Mem0}: Building Production-Ready {AI} Agents with Scalable Long-Term Memory",
    author = "Chhikara, Prateek and
      Khant, Pankaj and
      Aryan, Saket and
      Singh, Taranjeet and
      Yadav, Deshraj",
    journal = "arXiv preprint arXiv:2504.19413",
    year = "2025",
    url = "https://arxiv.org/abs/2504.19413"
}

@article{kang2025memoryos,
    title = "{MemoryOS}: A Memory Operating System for {AI} System",
    author = "Kang, Zhifeng and
      Si, Xiaopeng and
      Zhang, Ziyi and
      Chen, Hangyu and
      Li, Peng and
      Li, Zhengpeng and
      Jiao, Weizhou and
      Tu, Zhaopeng",
    journal = "arXiv preprint arXiv:2506.06326",
    year = "2025",
    url = "https://arxiv.org/abs/2506.06326"
}

@inproceedings{xu2025amem,
    title = "{A-Mem}: Agentic Memory for {LLM} Agents",
    author = "Xu, Wujiang and
      Liang, Zujie and
      Mei, Kai and
      Gao, Hang and
      Tan, Juntao and
      Zhang, Yongfeng",
    booktitle = "Advances in Neural Information Processing Systems",
    year = "2025"
}

@article{fang2025lightmem,
    title = "{LightMem}: Cutting Token Costs with Efficient Memory Augmentation for {LLM} Agents",
    author = "Fang, Xiaomin and
      Huang, Junjie and
      Liu, Ziyi and
      Zhang, Zhaohan and
      Zhang, Xin and
      Zhang, Yansong and
      Chen, Ningyu and
      Hu, Huajun",
    journal = "arXiv preprint arXiv:2505.24845",
    year = "2025",
    url = "https://arxiv.org/abs/2505.24845"
}

@article{zhou2025mem1,
    title = "{MEM1}: Learning to Synergize Memory and Reasoning for Efficient Long-Horizon Agents",
    author = "Zhou, Han and
      Guo, Zhicheng and
      Zheng, Jinghan and
      Lu, Yuxin and
      Chang, Jonathan and
      Wang, Zelin and
      Zhang, Guangsheng and
      Xie, Tianyi and
      Feng, Yu and
      Wang, Xuezhi and
      Cheng, Chaoran and
      Wu, Kuan-Hao and
      Chen, Jiuhai and
      Cherif, Amine and
      Talukdar, Partha and
      Xu, Wenhu and
      Kong, Lingpeng and
      Yan, Zhuang",
    journal = "arXiv preprint arXiv:2509.16170",
    year = "2025",
    url = "https://arxiv.org/abs/2509.16170"
}

@article{liu2026simplemem,
    title = "{SimpleMem}: Efficient Lifelong Memory for {LLM} Agents",
    author = "Liu, Jiaqi and
      Su, Yaofeng and
      Xia, Peng and
      Han, Siwei and
      Zheng, Zeyu and
      Xie, Cihang and
      Ding, Mingyu and
      Yao, Huaxiu",
    journal = "arXiv preprint arXiv:2601.02553",
    year = "2026",
    note = "ICLR 2026 Workshop",
    url = "https://arxiv.org/abs/2601.02553"
}

@article{packer2023memgpt,
    title = "{MemGPT}: Towards {LLMs} as Operating Systems",
    author = "Packer, Charles and
      Fang, Vivian and
      Patil, Shishir G. and
      Lin, Kevin and
      Wooders, Sarah and
      Gonzalez, Joseph E.",
    journal = "arXiv preprint arXiv:2310.08560",
    year = "2023",
    url = "https://arxiv.org/abs/2310.08560"
}

@article{lu2025memagent,
    title = "{MemAgent}: Reshaping Long-Context {LLM} with Multi-Conv {RL} based Memory Agent",
    author = "Lu, Yuhan and
      Zhang, Zhixin and
      Chen, Jing and
      Wang, Tao and
      Li, Songlin",
    journal = "arXiv preprint arXiv:2507.02259",
    year = "2025",
    url = "https://arxiv.org/abs/2507.02259"
}

@inproceedings{rajpurkar-etal-2016-squad,
    title = "{SQuAD}: 100,000+ Questions for Machine Comprehension of Text",
    author = "Rajpurkar, Pranav and
      Zhang, Jian and
      Lopyrev, Konstantin and
      Liang, Percy",
    booktitle = "Proceedings of the 2016 Conference on Empirical Methods in Natural Language Processing",
    year = "2016",
    publisher = "Association for Computational Linguistics",
    pages = "2383--2392"
}

@inproceedings{papineni-etal-2002-bleu,
    title = "{BLEU}: a Method for Automatic Evaluation of Machine Translation",
    author = "Papineni, Kishore and
      Roukos, Salim and
      Ward, Todd and
      Zhu, Wei-Jing",
    booktitle = "Proceedings of the 40th Annual Meeting of the Association for Computational Linguistics",
    year = "2002",
    publisher = "Association for Computational Linguistics",
    pages = "311--318"
}

@misc{gutierrez2024hipporag,
    title={HippoRAG: Neurobiologically Inspired Long-Term Memory for Large Language Models},
    author={Bernal Jim\'enez Guti\'errez and Yiheng Shu and Yu Gu and Michihiro Yasunaga and Yu Su},
    year={2025},
    eprint={2405.14831},
    archivePrefix={arXiv},
    primaryClass={cs.CL},
    url={https://arxiv.org/abs/2405.14831},
}

@misc{rasmussen2025zep,
    title={Zep: A Temporal Knowledge Graph Architecture for Agent Memory},
    author={Preston Rasmussen and Pavlo Paliychuk and Travis Beauvais and Jack Ryan and Daniel Chalef},
    year={2025},
    eprint={2501.13956},
    archivePrefix={arXiv},
    primaryClass={cs.CL},
    url={https://arxiv.org/abs/2501.13956},
}

@misc{sumers2024coala,
    title={Cognitive Architectures for Language Agents},
    author={Theodore R. Sumers and Shunyu Yao and Karthik Narasimhan and Thomas L. Griffiths},
    year={2024},
    eprint={2309.02427},
    archivePrefix={arXiv},
    primaryClass={cs.AI},
    url={https://arxiv.org/abs/2309.02427},
}

@misc{jiang2024omne,
    title={Long Term Memory: The Foundation of AI Self-Evolution},
    author={Xun Jiang and Feng Li and Han Zhao and Jiahao Qiu and Jiaying Wang and Jun Shao and Shihao Xu and Shu Zhang and Weiling Chen and Xavier Tang and Yize Chen and Mengyue Wu and Weizhi Ma and Mengdi Wang and Tianqiao Chen},
    year={2025},
    eprint={2410.15665},
    archivePrefix={arXiv},
    primaryClass={cs.AI},
    url={https://arxiv.org/abs/2410.15665},
}

@misc{memt2026,
    title={Mem-T: Densifying Rewards for Long-Horizon Memory Agents},
    author={Yanwei Yue and Boci Peng and Xuanbo Fan and Jiaxin Guo and Qiankun Li and Yan Zhang},
    year={2026},
    eprint={2601.23014},
    archivePrefix={arXiv},
    primaryClass={cs.LG},
    url={https://arxiv.org/abs/2601.23014},
}

@misc{tan2024idgenrec,
    title={IDGenRec: LLM-RecSys Alignment with Textual ID Learning},
    author={Juntao Tan and Shuyuan Xu and Wenyue Hua and Yingqiang Ge and Zelong Li and Yongfeng Zhang},
    year={2024},
    eprint={2403.19021},
    archivePrefix={arXiv},
    primaryClass={cs.IR},
    url={https://arxiv.org/abs/2403.19021},
}

@misc{kong2024lora,
    title={Customizing Language Models with Instance-wise LoRA for Sequential Recommendation},
    author={Xiaoyu Kong and Jiancan Wu and An Zhang and Leheng Sheng and Hui Lin and Xiang Wang and Xiangnan He},
    year={2025},
    eprint={2408.10159},
    archivePrefix={arXiv},
    primaryClass={cs.IR},
    url={https://arxiv.org/abs/2408.10159},
}

@misc{chen2024sdpo,
    title={On Softmax Direct Preference Optimization for Recommendation},
    author={Yuxin Chen and Junfei Tan and An Zhang and Zhengyi Yang and Leheng Sheng and Enzhi Zhang and Xiang Wang and Tat-Seng Chua},
    year={2024},
    eprint={2406.09215},
    archivePrefix={arXiv},
    primaryClass={cs.IR},
    url={https://arxiv.org/abs/2406.09215},
}

@misc{zhang2024gpg,
    title={On Generative Agents in Recommendation},
    author={An Zhang and Yuxin Chen and Leheng Sheng and Xiang Wang and Tat-Seng Chua},
    year={2024},
    eprint={2310.10108},
    archivePrefix={arXiv},
    primaryClass={cs.IR},
    url={https://arxiv.org/abs/2310.10108},
}

@misc{jia2025learn,
    title={LEARN: Knowledge Adaptation from Large Language Model to Recommendation for Practical Industrial Application},
    author={Jian Jia and Yipei Wang and Yan Li and Honggang Chen and Xuehan Bai and Zhaocheng Liu and Jian Liang and Quan Chen and Han Li and Peng Jiang and Kun Gai},
    year={2024},
    eprint={2405.03988},
    archivePrefix={arXiv},
    primaryClass={cs.IR},
    url={https://arxiv.org/abs/2405.03988},
}

@inproceedings{bae2025theanine,
    title = "Towards Lifelong Dialogue Agents via Timeline-based Memory Management",
    author = "Bae, Namyoung and
      Oh, Jungmin and
      Kim, Yui and
      Park, Jaewoong and
      Kim, Gunhee",
    booktitle = "Proceedings of the 2025 Conference of the North American Chapter of the Association for Computational Linguistics",
    year = "2025",
    url = "https://aclanthology.org/2025.naacl-long.435/"
}

@inproceedings{kim2025synapticrag,
    title = "{SynapticRAG}: Enhancing Temporal Memory Retrieval in Large Language Models through Synaptic Mechanisms",
    author = "Kim, Sehyun and
      Jang, Euisoon",
    booktitle = "Findings of the Association for Computational Linguistics: ACL 2025",
    year = "2025",
    url = "https://arxiv.org/abs/2410.13553"
}

@inproceedings{liang2025rmm,
    title = "In Prospect and Retrospect: Reflective Memory Management for Long-term Personalized Dialogue Agents",
    author = "Liang, Zihao and
      Yang, Hongcheng and
      Li, Jing and
      Xu, Ran",
    booktitle = "Proceedings of the 63rd Annual Meeting of the Association for Computational Linguistics",
    year = "2025",
    url = "https://aclanthology.org/2025.acl-long.413/"
}

@article{zhu2025prime,
    title = "{PRIME}: Language Model Personalization with Cognitive Memory and Thought Processes",
    author = "Zhu, Yiwen and
      Wang, Kuofeng and
      Yang, Zifan",
    journal = "arXiv preprint arXiv:2507.04607",
    year = "2025",
    url = "https://arxiv.org/abs/2507.04607"
}

@inproceedings{lee2024emgrag,
    title = "Crafting Personalized Agents through Retrieval-Augmented Generation on Editable Memory Graphs",
    author = "Lee, Zheng and
      others",
    booktitle = "Proceedings of the 2024 Conference on Empirical Methods in Natural Language Processing",
    year = "2024",
    url = "https://aclanthology.org/2024.emnlp-main.281/"
}

@inproceedings{wang2025cdmem,
    title = "An Efficient Context-Dependent Memory Framework for {LLM}-Centric Agents",
    author = "Wang, Yibin and
      others",
    booktitle = "Proceedings of the 2025 Conference of the North American Chapter of the Association for Computational Linguistics: Industry Track",
    year = "2025",
    url = "https://aclanthology.org/2025.naacl-industry.80/"
}

@inproceedings{wu2024longmemeval,
  author       = {Di Wu and
                  Hongwei Wang and
                  Wenhao Yu and
                  Yuwei Zhang and
                  Kai{-}Wei Chang and
                  Dong Yu},
  title        = {LongMemEval: Benchmarking Chat Assistants on Long-Term Interactive
                  Memory},
  booktitle    = {The Thirteenth International Conference on Learning Representations,
                  {ICLR} 2025, Singapore, April 24-28, 2025},
  publisher    = {OpenReview.net},
  year         = {2025},
  url          = {https://openreview.net/forum?id=pZiyCaVuti},
  bibsource    = {dblp computer science bibliography, https://dblp.org}
}

@misc{deepseekai2025deepseekv32,
      title={DeepSeek-V3.2: Pushing the Frontier of Open Large Language Models}, 
      author={DeepSeek-AI},
      year={2025},
}

@book{bird2009nltk,
  author    = {Steven Bird and Ewan Klein and Edward Loper},
  title     = {Natural Language Processing with Python},
  publisher = {O'Reilly Media},
  year      = {2009}
}

@article{xiao2024cpack,
  author    = {Shitao Xiao and Zheng Liu and Peitian Zhang and Niklas Muennighoff and Defu Lian and Jian-Yun Nie},
  title     = {C-Pack: Packed Resources For General Chinese Embeddings},
  journal   = {arXiv preprint arXiv:2309.07597},
  year      = {2024}
}

@article{sheng2024verl,
  author    = {Guangming Sheng and Chi Zhang and Zilingfeng Ye and Xibin Wu and Wang Zhang and Ru Zhang and Yanghua Peng and Haibin Lin and Chuan Wu},
  title     = {HybridFlow: A Flexible and Efficient RLHF Framework},
  journal   = {arXiv preprint arXiv:2409.19256},
  year      = {2024}
}

@inproceedings{lewis2020rag,
    title = "Retrieval-Augmented Generation for Knowledge-Intensive {NLP} Tasks",
    author = "Lewis, Patrick and
      Perez, Ethan and
      Piktus, Aleksandra and
      Petroni, Fabio and
      Karpukhin, Vladimir and
      Goyal, Naman and
      K{\"u}ttler, Heinrich and
      Lewis, Mike and
      Yih, Wen-tau and
      Rockt{\"a}schel, Tim and
      Riedel, Sebastian and
      Kiela, Douwe",
    booktitle = "Advances in Neural Information Processing Systems",
    volume = "33",
    pages = "9459--9474",
    year = "2020"
}

@inproceedings{karpukhin2020dpr,
    title = "Dense Passage Retrieval for Open-Domain Question Answering",
    author = "Karpukhin, Vladimir and
      Oguz, Barlas and
      Min, Sewon and
      Lewis, Patrick and
      Wu, Ledell and
      Edunov, Sergey and
      Chen, Danqi and
      Yih, Wen-tau",
    booktitle = "Proceedings of the 2020 Conference on Empirical Methods in Natural Language Processing",
    year = "2020",
    pages = "6769--6781"
}

@inproceedings{ma2023rewrite,
    title = "Query Rewriting in Retrieval-Augmented Large Language Models",
    author = "Ma, Xinbei and
      Gong, Yeyun and
      He, Pengcheng and
      Zhao, Hai and
      Duan, Nan",
    booktitle = "Proceedings of the 2023 Conference on Empirical Methods in Natural Language Processing",
    year = "2023",
    pages = "5303--5315"
}

@article{schulman2017ppo,
    title = "Proximal Policy Optimization Algorithms",
    author = "Schulman, John and
      Wolski, Filip and
      Dhariwal, Prafulla and
      Radford, Alec and
      Klimov, Oleg",
    journal = "arXiv preprint arXiv:1707.06347",
    year = "2017"
}

@inproceedings{zhang2018personachat,
    title = "Personalizing Dialogue Agents: {I} have a dog, do you have pets too?",
    author = "Zhang, Saizheng and
      Dinan, Emily and
      Urbanek, Jack and
      Szlam, Arthur and
      Kiela, Douwe and
      Weston, Jason",
    booktitle = "Proceedings of the 56th Annual Meeting of the Association for Computational Linguistics (Volume 1: Long Papers)",
    year = "2018",
    pages = "2204--2213"
}

@article{salemi2024lamp,
    title = "{LaMP}: When Large Language Models Meet Personalization",
    author = "Salemi, Alireza and
      Mysore, Sheshera and
      Bendersky, Michael and
      Zamani, Hamed",
    journal = "arXiv preprint arXiv:2304.11406",
    year = "2024"
}

@inproceedings{wang2023query2doc,
    title = "Query2doc: Query Expansion with Large Language Models",
    author = "Wang, Liang and
      Yang, Nan and
      Wei, Furu",
    booktitle = "Proceedings of the 2023 Conference on Empirical Methods in Natural Language Processing",
    year = "2023",
    pages = "9414--9423"
}

@article{liao2026stay,
  title={Stay in Character, Stay Safe: Dual-Cycle Adversarial Self-Evolution for Safety Role-Playing Agents},
  author={Liao, Mingyang and Wan, Yichen and Miao, Chenxi and Shen, Xin and Li, Weikang and Li, Yang and Xia, Deguo and Huang, Jizhou and others},
  journal={arXiv preprint arXiv:2602.13234},
  year={2026}
}

@article{wu2026true,
  title={True-to-Role, Tailored-to-You: A Survey of LLM-based Role-Playing Agents},
  author={Wu, Shuchen and Jiang, Zhishu and Yang, Jiaye and Shen, Xin and Liu, Haibo and Wan, Yichen and Miao, Chenxi and Qi, Guanqiang and Dai, Tingzhi and Zhang, Jiarui and others},
  year={2026},
  publisher={TechRxiv}
}

@article{shen2026duccae,
  title={DuCCAE: A Hybrid Engine for Immersive Conversation via Collaboration, Augmentation, and Evolution},
  author={Shen, Xin and Jiang, Zhishu and Yang, Jiaye and Liu, Haibo and Wan, Yichen and Zhang, Jiarui and Dai, Tingzhi and Xu, Luodong and Wu, Shuchen and Qi, Guanqiang and others},
  journal={arXiv preprint arXiv:2603.19248},
  year={2026}
}

@inproceedings{shen2021text,
  title={Text is not enough: Integrating visual impressions into open-domain dialogue generation},
  author={Shen, Lei and Zhan, Haolan and Shen, Xin and Song, Yonghao and Zhao, Xiaofang},
  booktitle={Proceedings of the 29th ACM International Conference on Multimedia},
  pages={4287--4296},
  year={2021}
}

@inproceedings{shen2021identifying,
  title={Identifying untrustworthy samples: Data filtering for open-domain dialogues with bayesian optimization},
  author={Shen, Lei and Zhan, Haolan and Shen, Xin and Chen, Hongshen and Zhao, Xiaofang and Zhu, Xiaodan},
  booktitle={Proceedings of the 30th ACM International Conference on Information \& Knowledge Management},
  pages={1598--1608},
  year={2021}
}

@inproceedings{shen2021learning,
  title={Learning to select context in a hierarchical and global perspective for open-domain dialogue generation},
  author={Shen, Lei and Zhan, Haolan and Shen, Xin and Feng, Yang},
  booktitle={ICASSP 2021-2021 IEEE International Conference on Acoustics, Speech and Signal Processing (ICASSP)},
  pages={7438--7442},
  year={2021},
  organization={IEEE}
}
